\documentclass[aps,pra,amsmath,amssymb,superscriptaddress,reprint]{revtex4-1}

\usepackage{bm}
\usepackage{multirow}
\usepackage{verbatim}
\usepackage{longtable}
\usepackage{hyperref}
\usepackage[usenames,dvipsnames]{xcolor}

\usepackage{graphicx}
\usepackage{float}
\usepackage{color}

\begin{document}
\title{Scalable integrated single-photon source}
\author{Ravitej Uppu}
\email{ravitej.uppu@nbi.ku.dk}
\affiliation{Center for Hybrid Quantum Networks (Hy-Q), Niels Bohr Institute, University of Copenhagen, Blegdamsvej 17, DK-2100 Copenhagen, Denmark}
\author{Freja T. Pedersen}
\affiliation{Center for Hybrid Quantum Networks (Hy-Q), Niels Bohr Institute, University of Copenhagen, Blegdamsvej 17, DK-2100 Copenhagen, Denmark}
\author{Ying Wang}
\affiliation{Center for Hybrid Quantum Networks (Hy-Q), Niels Bohr Institute, University of Copenhagen, Blegdamsvej 17, DK-2100 Copenhagen, Denmark}
\author{Cecilie T. Olesen}
\affiliation{Center for Hybrid Quantum Networks (Hy-Q), Niels Bohr Institute, University of Copenhagen, Blegdamsvej 17, DK-2100 Copenhagen, Denmark}
\author{Camille Papon}
\affiliation{Center for Hybrid Quantum Networks (Hy-Q), Niels Bohr Institute, University of Copenhagen, Blegdamsvej 17, DK-2100 Copenhagen, Denmark}
\author{Xiaoyan Zhou}
\affiliation{Center for Hybrid Quantum Networks (Hy-Q), Niels Bohr Institute, University of Copenhagen, Blegdamsvej 17, DK-2100 Copenhagen, Denmark}
\author{Leonardo Midolo}
\affiliation{Center for Hybrid Quantum Networks (Hy-Q), Niels Bohr Institute, University of Copenhagen, Blegdamsvej 17, DK-2100 Copenhagen, Denmark}
\author{Sven Scholz}
\affiliation{Lehrstuhl f{\"u}r Angewandte Festk{\"o}rperphysik, Ruhr-Universit{\"a}t Bochum, Universit{\"a}tsstrasse 150, D-44780 Bochum, Germany}
\author{Andreas D.~Wieck}
\affiliation{Lehrstuhl f{\"u}r Angewandte Festk{\"o}rperphysik, Ruhr-Universit{\"a}t Bochum, Universit{\"a}tsstrasse 150, D-44780 Bochum, Germany}
\author{Arne Ludwig}
\affiliation{Lehrstuhl f{\"u}r Angewandte Festk{\"o}rperphysik, Ruhr-Universit{\"a}t Bochum, Universit{\"a}tsstrasse 150, D-44780 Bochum, Germany}
\author{Peter Lodahl}
\email{lodahl@nbi.ku.dk}
\affiliation{Center for Hybrid Quantum Networks (Hy-Q), Niels Bohr Institute, University of Copenhagen, Blegdamsvej 17, DK-2100 Copenhagen, Denmark}

\date{\today}

\begin{abstract}
Photonic qubits are key enablers for quantum-information processing deployable across a distributed quantum network. An on-demand and truly scalable source of indistinguishable single photons is the essential component enabling high-fidelity photonic quantum operations. A main challenge is to overcome noise and decoherence processes in order to reach the steep benchmarks on generation efficiency and photon indistinguishability required for scaling up the source. We report on the realization of a deterministic single-photon source featuring near-unity indistinguishability using a quantum dot in an `on-chip' planar nanophotonic waveguide circuit. The device produces long strings of $>100$ single photons without any observable decrease in the mutual indistinguishability between photons. A total generation rate of 122 million photons per second is achieved corresponding to an `on-chip' source efficiency of $84 \%$. These specifications of the single-photon source are benchmarked for boson sampling and found to enable scaling into the regime of quantum advantage.
\end{abstract}
\maketitle

Leveraging photonic quantum technology requires scalable hardware.
A key enabling device is a high-quality and on-demand source of indistinguishable single photons with immediate use for quantum simulators \cite{walther2012}, device-independent quantum communication \cite{acin2018}, memoryless quantum repeaters \cite{Borregaard2019}, or as a primer for multi-photon entanglement sources \cite{Lindner2009}. Furthermore, single photons are the natural carriers of quantum information over extended distances, thereby providing a backbone for the quantum internet \cite{kimble2008} by enabling fully-secure quantum communication \cite{hanson2018} and a modular approach to quantum computing \cite{Rudolph2017,monroe2013}.

An on-demand source of indistinguishable single photons is the major building block that can be realized either with a probabilistic source, which can be heralded and multiplexed to improve efficiency \cite{kwiat2019}, or using a single quantum emitter coupled to a waveguide or cavity designed to collect the spontaneously emitted single photons.
Significant progress has been made with the latter approach by coupling quantum dots (QDs) to photonic nanostructures\cite{lodahl2014,lodahl2015,senellart2016,hofling2016,pan2017,hofling2017,lodahl2017,fox2018,pan2019elliptical,lodahl2018,Uppu2020}, and the governing fundamental processes determining performance have now clearly been identified including decoherence processes \cite{lodahl2017quantum}.
Nonetheless, deterministic operation of a source of high-quality indistinguishable photons on a scalable platform has not yet been achieved, which is a key enabling step towards demonstrating quantum advantage with single photons
\cite{preskill2018,Aaronson2016a}. Quantum advantage has so far been reported with superconducting qubits \cite{Arute2019} while state-of-the-art in photonics is the 20-photon experiment reported with a QD source \cite{pan2019boson}.
Deterministic and coherent operation requires a number of simultaneous capabilities: i) the QD must be deterministically and resonantly excited with a tailored optical pulse whilst eliminating the excess pump light without reducing the single-photon purity and efficiency, ii) the emitted photon must be efficiently coupled to a single propagating mode, iii) electrical control of the QD must be implemented to overcome efficiency loss due to emission into other QD charge states, and iv) decoherence and noise processes must be eliminated over the relevant time scale \cite{kuhlmann2013} in order to produce a scalable source of multiple indistinguishable photons .

In the present work, we implement all four functionalities in a single device, using a QD efficiently coupled to an electrically contacted planar photonic-crystal waveguide membrane.
We generate temporal strings of $>100$ single photons with pairwise photon indistinguishability exceeding $96\%$.
Such a source coupled with an active temporal-to-spatial mode demultiplexer \cite{pan2017,lodahl2019demux}, will set new standards for multi-photon experiments aimed at establishing photonic quantum advantage that requires more than 50 photons \cite{ralph2017,preskill2018}.
In a recent breakthrough experiment, up to 20 photons were employed in a boson sampling experiment \cite{pan2019boson}, however the photon indistinguishability was observed to decay over the 20-photon chain.
This loss of coherence was also reported in previous experiments and only explained heuristically \cite{pan2016}, and is likely a consequence of insufficient control of the QD charge environment leading to noise.
In our improved source, we demonstrate coherence extending to at least 115 photons, as is proven by measuring the mutual degree of indistinguishability between photons emitted with a time delay approaching a microsecond.
The source efficiency specifying the `on-chip' generation probability of indistinguishable single photons is $\eta_S = 84\%$ comprising of $92\%$ coupling of the dipole to the waveguide (the $\beta$-factor), $95\%$ efficient emission into the coherent zero-photon line, $98\%$ radiative decay efficiency, and $>98\%$ emission of the best-coupled of the two linear dipoles.
Overall we demonstrate the generation of 122 million photons per second in the waveguide, which is the main efficiency figure-of-merit that the present experiment targets. 
This massive photonic quantum resource is coupled off-chip and into an optical fiber, with an efficiency limited only by minor residual loss ($4 \%$) in the waveguide and a chip-to-fiber outcoupling efficiency that reaches up to $82 \%$ with optimized grating outcouplers. The full details of the efficiency characterizations are presented in the Appendix in addition with an account of how to optimize external coupling efficiencies with already demonstrated methods in order to reach a fiber-coupled source with an overall efficiency of $78 \%$. The improved source coherence will result in shorter runtimes for the validation of boson sampling in the quantum regime, thereby overcoming a major technological challenge.
Our work lays a clear path way for demonstrating quantum advantage in boson sampling with $>50$ photons using the source in combination with realistic low-loss optical networks and high-efficiency detectors.

\begin{figure}
\centering
\includegraphics{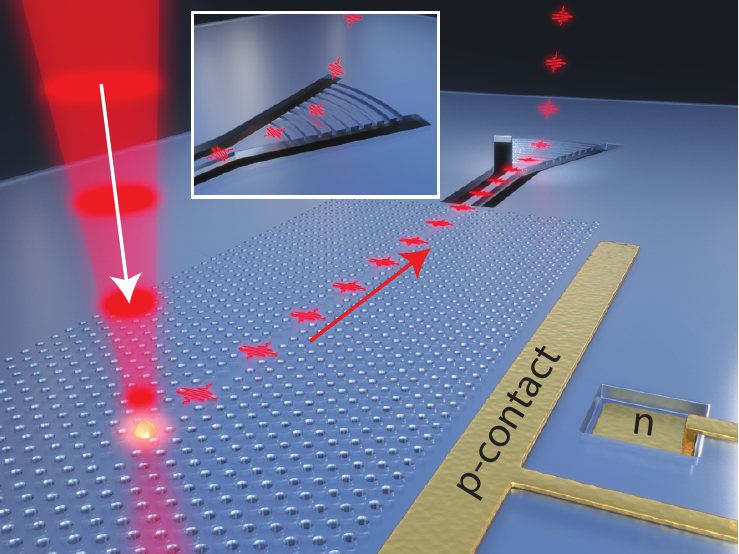}
\caption{Illustration of the single-photon source device. A QD embedded in a photonic crystal waveguide is excited using a pulsed laser at the resonance wavelength of the QD. The emitted single-photon train is coupled to the waveguide with near-unity efficiency and outcoupled from the device using a grating outcoupler (see inset). Metal electrical contacts (shown in gold) are used for applying a gate voltage across the QD embedded in the 170 nm thin membrane.}
\label{fig1}
\end{figure}

\begin{figure*}
\centering
\includegraphics[width=\textwidth]{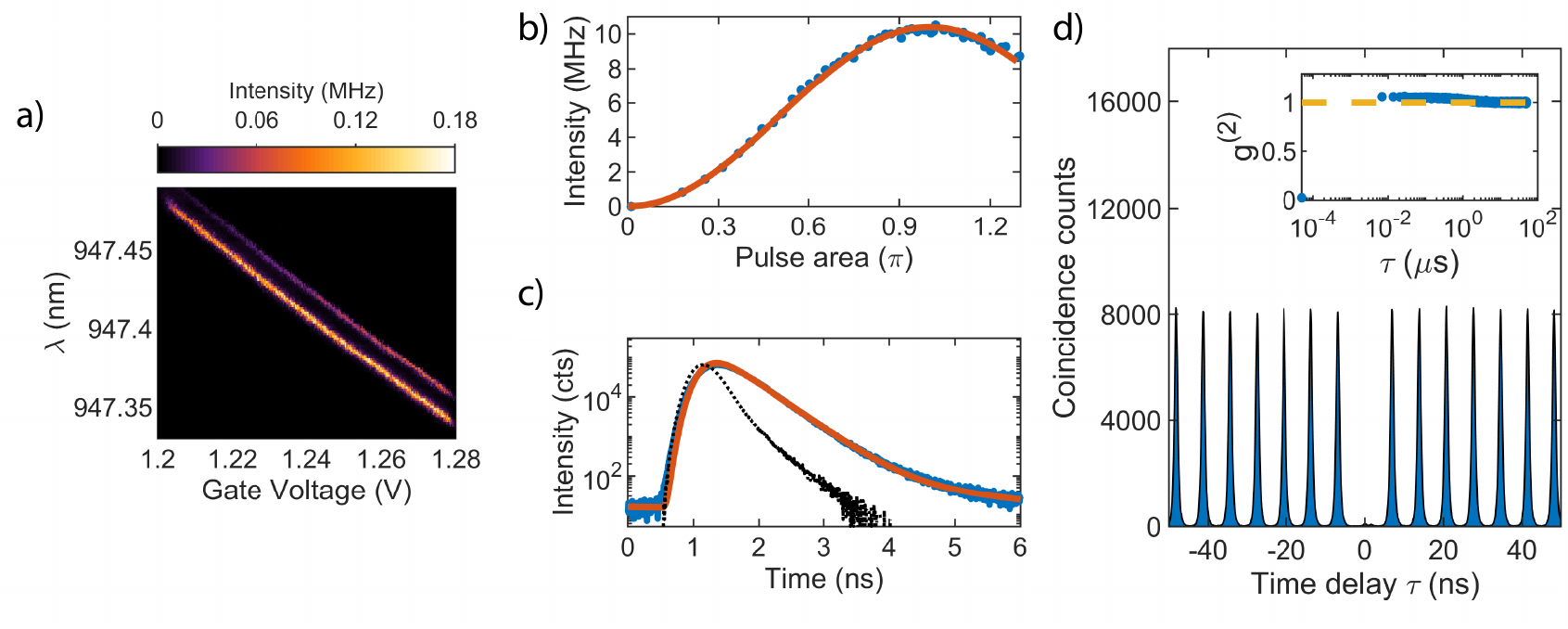}
\caption{ Deterministic preparation of an excitation in the QD. (a) Resonance fluorescence measured from a QD in a photonic-crystal waveguide weakly excited using a continuous wave tunable diode laser. The two bright lines are the charge plateaus of the fine structure split neutral exciton. (b) Pulsed resonance fluorescence measured with the QD tuned on resonance ($V_g = 1.24$ V) and excited with a mode-locked laser with pulse width of 20 ps. (c) Lifetime of the resonantly excited QD exhibiting a single exponential decay with a radiative decay rate of $\gamma = 2.89$ ns$^{-1}$. The black dotted curve is the instrument response function of the measurement setup. (d) Measured coincidence counts of the single-photon source in a Hanbury Brown and Twiss interferometer under $\pi$-pulse excitation showing a strongly suppressed peak at time delay $\tau$ = 0. The inset shows the integrated coincidence counts under each peak over a timescale of 50 $\mu$s that highlights the minimal bunching observed.}
\label{fig2}
\end{figure*}

\emph{Operational principle of the single-photon device:} Figure \ref{fig1} displays the device comprising epitaxially grown QDs embedded in a 170 nm thin membrane (see Appendix A for details on sample fabrication).
The QD is excited with short optical pulses whereby an excitation in the QD can be deterministically prepared.
The emitted single photons are channeled on-demand into a photonic-crystal waveguide designed to control the local density of optical states such that an embedded QD emits with near-unity coupling efficiency (quantified by the $\beta$-factor) into the waveguide \cite{lodahl2014}.
The collected photons are subsequently routed on-chip and directed to a tailored grating for highly-efficient outcoupling to an optical fiber.
The spatial separation between the excitation laser and the collection grating ensures that very high suppression of the pulsed resonant laser can be obtained without employing any polarization filtering that could result in losses.
Figure \ref{fig2}(b) shows an example of pulsed resonance-fluorescence data that exhibit clear Rabi oscillation with highly suppressed laser background.

\emph{Demonstration of low-noise operation:}
The implementation of electrical contacts on the device, cf. Fig. \ref{fig1}, leads to a number of salient features: The embedded QDs can be electrically tuned, the charge state of the QD is stabilized so that recombination only on the desired transition takes place, and spectral diffusion due to residual charge noise in the structures can be strongly suppressed.
As a consequence,  near-transform-limited optical linewidths can be achieved in the photonic nanostructures \cite{kuhlmann2015,lodahl2018}, which is essential for generating a scalable resource of indistinguishable photons as well as for more advanced applications of the system for photonic quantum gates and entanglement generation \cite{lodahl2017quantum}.
Low-noise operation is demonstrated by exciting the QD with a tunable laser and collecting the resonance fluorescence.
A typical measurement is displayed in Fig. \ref{fig2}(a), where two distinct QD transitions (the two orthogonal dipoles of the neutral exciton) are visible and the excitation laser is clearly suppressed.
A distinct Coulomb-blockade regime is observed \cite{warburton2013} meaning that the QD emits solely on the identified neutral exciton transition, i.e. blinking to other exciton complexes is fully suppressed.
We observe a QD linewidth of $\approx$800 MHz, which is close to the transform limit, and the slight residual broadening is attributed to slow-time drift (1--10 ms) \cite{srinivasan2018}, which is irrelevant for the generation of indistinguishable photons over the nano-micro second time scales studied here.

\emph{Deterministic single-photon generation:}
Pulsed resonant excitation allows on-demand operation of the single-photon device.
The QD was excited with $20$ ps pulses and clear Rabi oscillations are observed when increasing the excitation power, cf. Fig. \ref{fig2}(b).
Deterministic operation corresponds to excitation with a $\pi-$pulse, where essentially back-ground free operation is observed with a very low single-photon impurity contribution of $\xi <0.007$.
Here $\xi$ is defined as the ratio of the laser background to the QD signal intensity.
The single-photon purity is quantified in second-order photon correlation measurements, cf. data in Fig.\ref{fig2}(d).
We extract $g^{(2)}(0) = 0.015 \pm 0.005$, which can be further improved by engineering the resonant excitation pulse \cite{Sumanta2019} or by implementing two-photon excitation schemes \cite{zwiller2018}. Importantly, blinking of the source is essentially vanishing (cf. inset of Fig. \ref{fig2}(d)) up to time-scales approaching milliseconds (data up to $50 \mu$s shown).
The photon emission dynamics is reproduced in Fig. \ref{fig2}(c) where a radiative decay rate of $\gamma = 2.89$ ns$^{-1}$ is extracted for the efficiently coupled dipole, which is enhanced by the Purcell effect of the waveguide leading to the large $\beta$-factor \cite{lodahl2015}.

\begin{figure*}
\centering
\includegraphics[width=\textwidth]{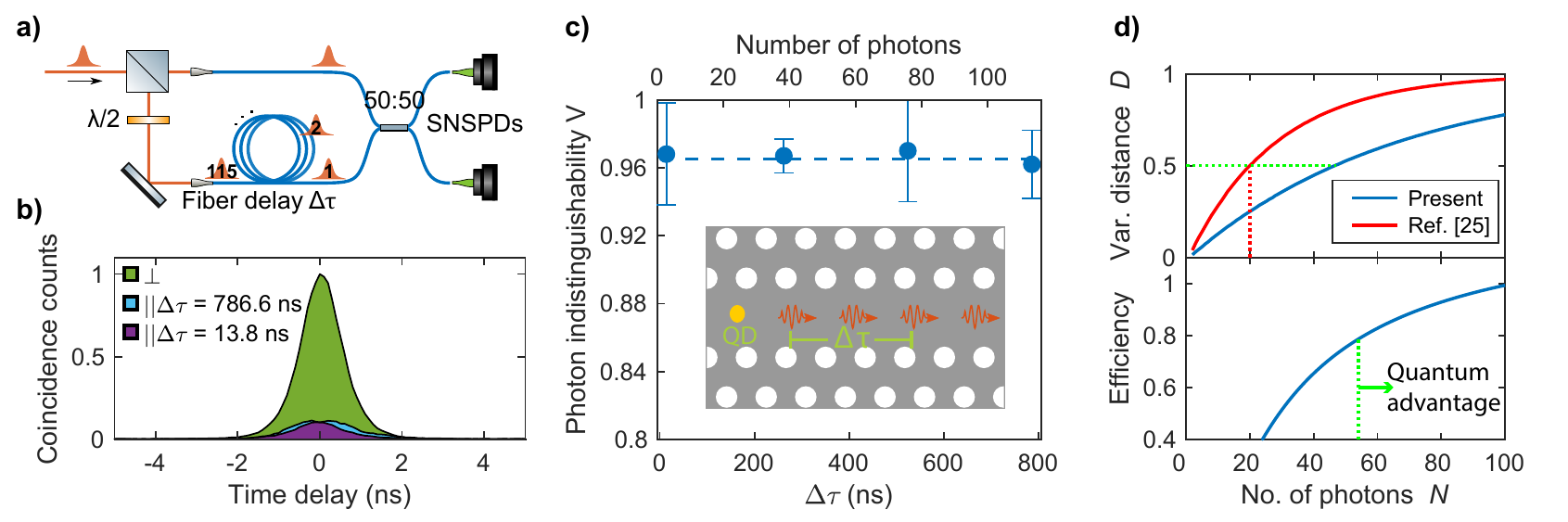}
\caption{ Highly-indistinguishable train of single photons. (a) Schematic of a fiber-based unbalanced Mach-Zehnder interferometer with a variable fiber delay line (delay time $\Delta \tau$) in one arm employed for HOM interference measurements. (b) Photon indistinguishability measured under $\pi-$pulse excitation for photons generated with $\Delta \tau = 13.8$ ns and $\Delta\tau = 786.6$ ns, and with the two input photons co- and cross-polarized by adjusting the half-wave plate in the fiber-delay arm. (c) Photon indistinguishability between photon pairs in a temporal string of up to 115 photons reaching $>96\%$, as illustrated by the interference of photon 1 with photons 2, 39, 77, and 115. (d) Estimate of boson sampling capabilities of QD sources. Top panel: variational distance of an ideal boson sampler from the real scenario implemented with the present source (blue curve) and that from Ref. \cite{pan2019boson} (red curve). At a given $N$, higher trace distance requires more sampling events and hence longer time to validate the boson sampler, thereby inhibiting the scaling into quantum advantage. Bottom panel: minimum source efficiency $\eta_S$ required for validating boson sampling with $N$ indistinguishable photons by detecting collision-free events in a fixed runtime of 30 days.}
\label{fig3}
\end{figure*}

\emph{Generation of long strings of indistinguishable photons:} The indistinguishability of the temporal single photon train is quantified through photon-photon interference experiments.
In these measurements two photons emitted at different times are interfered in an asymmetric interferometer with a variable time delay, as schematized in Fig. \ref{fig3}(a).
In this setup, we can measure the degree of indistinguishability between single photons emitted from the QD with time intervals $N\tau_p$, where $\tau_p$ is the laser repetition period and $N$ is a positive integer.
Figure \ref{fig3}(c) shows experimental data for the four representative values  $N = \{1,38,76,114\}$, where the latter corresponds to  a maximum time delay between two photons of $786.6$ ns.
Figure \ref{fig3}(b) shows the recorded correlation histograms for $N = 1$ and $N=114$ for the two cases where the interfering photons are co- and cross-polarized.
The cross-polarized histogram serves as a reference measurement for extracting the degree of indistinguishability $V$ after accounting for the setup imperfections (cf. Appendix C for the analysis).
We find $V = (96 \pm 2)\%$ when accounting for the finite multi-photon probability discussed above, while $V = (93 \pm 2)\%$ is directly recorded. 
The minor amount of  distinguishability can be attributed to residual phonon decoherence \cite{sorensen2018}, which is the fundamental decoherence process that determines the performance of QD single-photon sources. 
Importantly, we find the photon indistinguishability to remain over $96 \%$ for delays corresponding to 115 photons, cf. Fig. \ref{fig3}(c), which is key for the applicability of the source to reach quantum advantage, as benchmarked below.

\emph{Benchmarking quantum advantage with the source:}
Scalable boson sampling experiments employing QD sources utilize active switching of the temporal string of single photons into separate optical modes, thereby realizing a multiphoton source.
Even a small degree of distinguishability of photons can strongly influence the scalability of boson sampling into the quantum advantange regime \cite{Renema2018}.
The impact of the improved source coherence on boson sampling can be quantified using the variational distance $D$ of boson sampling, which is the statistical distance between the probability distributions of photon correlations, implemented using the partially distinguishable photon source and an ideal source \cite{Tichy2015,Shchesnovich2015}.
For distinguishable photons in a Haar unitary optical network,  $D \approx 1$ for large $N$.
Therefore, validation of boson sampling against the classically simulatable case of distinguishable photons requires $D<1$; with larger $D$ demanding higher number of multiphoton detection events.
Better source coherence, i.e. higher pair-wise indistinguishability across the string, results in a lower $D$ for any $N$-photon boson sampling, as shown in the top panel of Fig. \ref{fig3}(d).
Importantly, at the comparable $D$ as the 20-photon boson sampling in Ref. \cite{pan2019boson}, the better source coherence reported here enables boson sampling with $54$ photons, see Appendix E for details.
Figure \ref{fig3}(d) exploits the required efficiency of the source for boson sampling versus number of photons for a technologically feasible operation time of 30 days.
We find that reaching the regime of quantum advantage requires an efficiency $\eta >78\%$, which is met by the current on-chip source and even achievable with the projected fiber-coupled source, cf. discussion in Appendix D.4. We emphasize that near-unity indistinguishability extending over the whole string of generated photons, which is achieved in our low-noise devices, is essential for scaling-up to reach the quantum advantage threshold 
We note that given the high quality photonic resource, these runtimes could be further improved using the Aaronson-Brod model of boson sampling \cite{Aaronson2016}.

We have presented a scalable single-photon source based on a QD in a photonic waveguide meeting the very strict requirements needed for demonstrating quantum advantage of photonic qubits.
Reaching quantum advantage is a crucial first step towards advanced quantum simulators and computers that provides clear benchmarks on the metrics for quantum hardware.
Notably these benchmarks are universal, i.e. they are also essential figures-of-merit for more advanced photonic quantum resources produced with QD sources than the independent photons required for boson sampling.
Consequently, our work is expected to spur significant interest towards application of QD sources for deterministic photonic quantum gates, multi-photon entanglement generation, and nonlinear quantum optics.

\appendix
\section{Heterostructure composition and sample fabrication}
\begin{figure*}
\centering
\includegraphics[width=\textwidth]{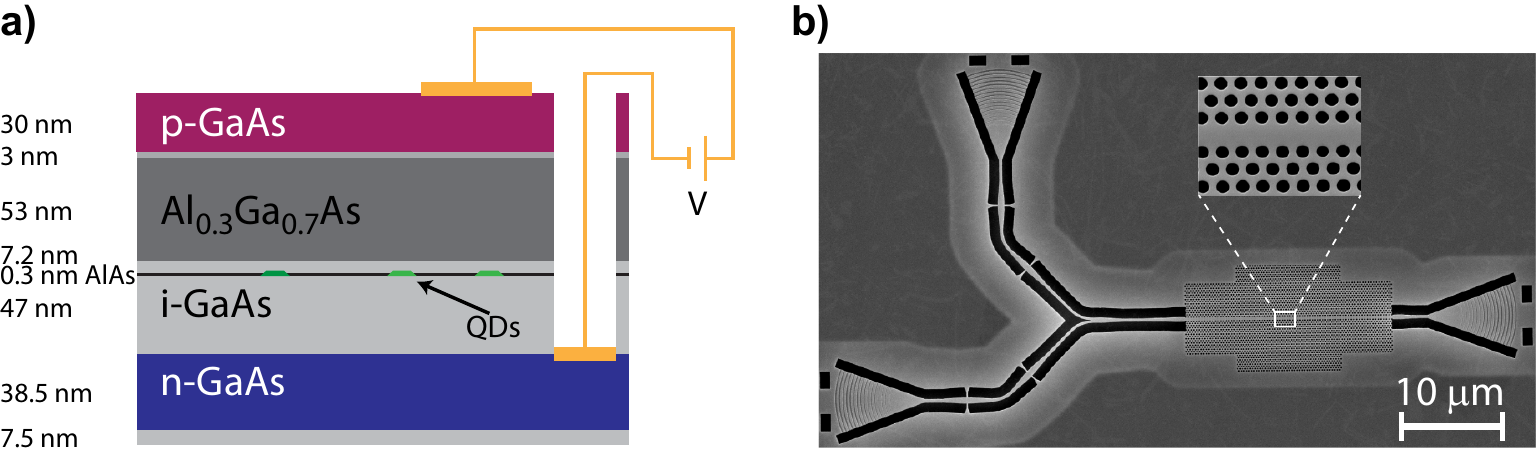}
\caption{ (a) Outline of $p$-$i$-$n$ diode heterostructure used to realize the device. (b) Scanning electron microscope image of the photonic crystal waveguide employed in the measurements with lattice constant $a = 248$ nm and hole radius $r = 70$ nm. The photonic crystal waveguides are terminated with high-efficiency shallow-etched grating outcouplers for collecting the QD emission.}
\label{figS1}
\end{figure*}
The samples are fabricated on a GaAs membrane grown by molecular beam epitaxy on a (100) GaAs substrate.
The substrate is prepared for growth using an AlAs/GaAs superlattice followed by the growth of a 1150-nm-thick Al$_{0.75}$Ga$_{0.25}$As sacrificial layer.
Subsequently, a 170 nm-thin GaAs membrane containing InAs quantum dots (QDs) is grown on top of the sacrifical layer.
The membrane constitutes an ultra-thin $p$-$i$-$n$ diode with the heterostructure shown in Fig. \ref{figS1}(a).
The $p$-$i$-$n$ diode is used to apply an electric field across the QDs to reduce charge noise and control the charge state as well as Stark tune the QD emission wavelength.
The epitaxial $n$- and $p$-type regions are realized by doping the GaAs during the growth with silicon and carbon, respectively.
The layer of self-assembled InAs QDs is located at the center of the membrane in order to maximally couple to the optically guided TE-mode.
The $n$-type region is located at a distance of 47 nm from the QDs to ensure the suppression of cotunneling.
The QDs are capped with a single monolayer of AlAs that assists in removing the electron wetting layer states \cite{Lobl2019}.
A 53-nm-thick Al$_{0.3}$Ga$_{0.7}$As layer above the QDs is used as a tunnel barrier to limit the current to a few nA at a bias voltage of $\approx$ 1 V, where the QDs can be charged with a single electron.

The first step in creating the nanostructures is the fabrication of the electrical contacts to the $p$-doped and $n$-doped layers.
Reactive-ion etching (RIE) in a BCl$_3$/Ar chemistry is used to open vias to the $n$-layer and Ni/Ge/Au/Ni/Au contacts are deposited by electron-beam physical vapor deposition and annealed at 430 $^\circ$C.
Subsequently Cr/Au pads are deposited on the surface to realize Ohmic $p$-type contacts.
The chip of size $3$ mm $\times$ $3$ mm is divided into five sections with physical dimensions of $2.5$ mm $\times$ $0.5$ mm.
Each of these sections is connected to separate pairs of electrical contacts.
In order to achieve minimum cross-talk between the different sections, an isolation trench with a width of 1 $\mu$m is patterned.
The nanostructuring is then carried out using a soft-mask-based process described in Ref. \cite{midolo_soft_2015}.
The shallow-etched grating outcouplers are patterned by electron-beam lithography (Elionix F-125; 125 keV electron beam) and then etched using reactive ion etching (RIE) together with the isolation trenches aimed at shortening the fabrication process.
The photonic crystal waveguides are subsequently patterned using another electron beam lithography step.
These patterns are then etched in the GaAs membrane using an inductively-coupled plasma reactive ion etching (ICP/RIE) in a BCl$_3$/Cl$_2$/Ar chemistry.
The residual polymer from the soft mask is removed by dipping the sample in N-Methyl-pyrrolidone at 70 $^\circ$C to the sample after the ICP/RIE process.
The sacrificial layer is then removed using wet etching using hydrofluoric acid to release the membrane.

A scanning electron microscope (SEM) image of one of the photonic crystal waveguides with lattice constant $a = 248$ nm and hole radius $r$ = 70 nm on the processed chip is shown in Fig. \ref{figS1}(b).
The bidirectional waveguides with grating outcouplers at each end is used to perform resonant transmission measurement for selecting QDs that are well-coupled to the photonic crystal waveguide.
The Y-splitter in one of the arms can be employed to perform on-chip Hanbury Brown and Twiss measurements by collecting the QD emission from the two gratings.


\section{Experimental setup}
The sample is cooled to 1.6 K in a cryostat with optical and electrical access.
The QD is excited from the top of the chip using a wide field-of-view confocal microscope with a high numerical aperture objective (NA = 0.81); see Fig. \ref{figS2}(a).
The emission from the QD is collected at the grating outcoupler through the same objective and imaged onto a single-mode optical fibre.
The excitation and collection paths are separated at a 5:95 (reflection:transmission) beam splitter, with the $95\%$ transmission path for collection.
A set of quarter and half wave plates in the excitation path allow polarization control of the excitation laser.
The QD emission collected in the fibre is spectrally filtered using a $3$ GHz linewidth etalon (free spectral range: $100$ GHz) to suppress the emission in the phonon sideband.
The intensity of the spectrally filtered emission is measured using a fibre-coupled superconducting nanowire single photon detector (SNSPD).
The bias voltage across the QD is tuned using a low-noise high-resolution DC voltage source.

\section{Analyzing the indistinguishability data}
The measured visibility of the HOM interference is affected by 1) the symmetry of the beam splitter employed in the interferometer (where $R$ and $T$ are the beam splitter's reflectance and transmittance, respectively) and 2) the precision in aligning the interferometer for maximal classical interference .
The former is limited by the beam splitter reflectivity, which in our setup was calibrated to be $R = 0.476$ and $T = 0.524$.
The latter is maximized in a best-effort approach at the beginning of the measurement to a typical value of $(1-\epsilon) \approx 0.995$.
We employ the procedure discussed in Ref.~\cite{fox2018} and correct the raw indistinguishability for setup imperfections and finite $g^{(2)}(0)$.

\begin{figure}
\centering
\includegraphics[width=\columnwidth]{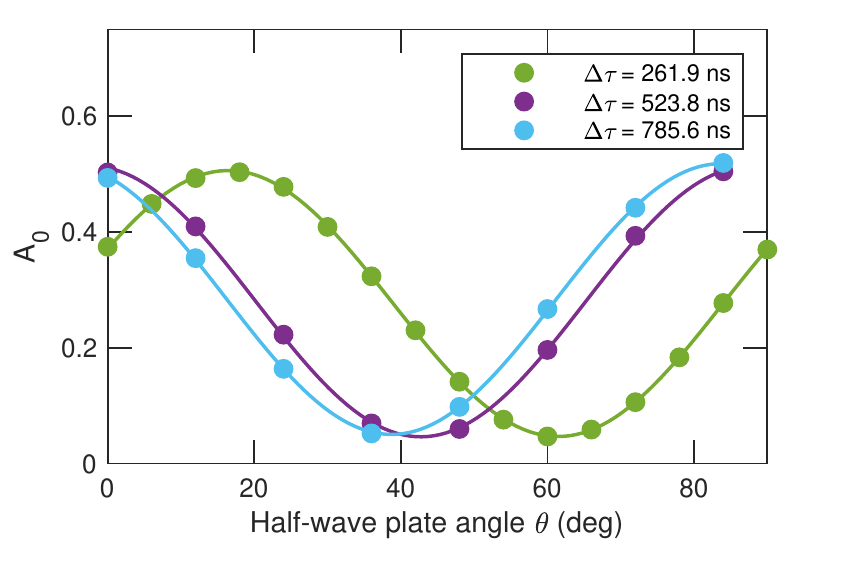}
\caption{Amplitude of the peak at zero delay $A_0$ of the correlation histogram for varying polarisation mismatch between the two photons. The polarisation was controlled by rotating the angle of a half-wave plate $\theta$. The fit to the data yields the measured visibility $V_\textrm{raw}$ without accounting for setup imperfections and finite $g^{(2)} (0)$. The dependence of $A_0$ on $\theta$ is measured for three fiber delays $\Delta \tau$, corresponding to interference between photons separated by $\Delta \tau$ in the temporal string. The offset in $\theta$ is chosen to offset the data for clarity.}
\label{figS6}
\end{figure}

The peaks in the measured coincidence counts (c.f. Fig. \ref{fig3}(b)) are fitted with two-sided single exponential decays convoluted with the instrument response function of the single photon detectors.
The correlation histogram is normalized to the amplitude of the peak at long time delay $A_{(t = 50 \mu s)}$\cite{lodahl2017}.
This procedure of measuring, fitting, and normalizing the correlation histograms is carried out for varying polarisation mismatch of the photons incident at the beam splitter.
The polarisation mismatch is varied by rotating the half-wave plate $\lambda/2$ in the fiber delay arm (c.f. Fig. \ref{fig3}(a)) between 0 and 90 degrees.
The normalization procedure removes the dependence on the total number of detected coincidences over the measurement time required for the polarisation scan.
Figure \ref{figS6} shows a plot of the normalized peak amplitude at zero time delay $A_0$ measured at each of the half-wave plate settings.
$A_0$ maximizes for perfectly distinguishable photons (cross-polarized) and minimizes when the photons are maximally indistinguishable (co-polarized).
The dependence of $A_0$ on the half-wave plate angle $\theta$ is fitted to the function
\begin{equation}
A_0 (\theta) = A_m - A_c \sin^2(2\theta + \phi),
\end{equation}
where, $\phi$ is a fitting factor that accounts for any offsets in the half-wave plate position. The amplitudes $A_m$ and $A_c$ are related to the measured HOM visibility $V_\textrm{raw} := A_c/A_m$ that does not account for setup imperfections.

From the measured raw visibility $V_\mathrm{raw}$, we can extract the intrinsic visibility $V$ by accounting for the slight imbalance of the measurement interferometer and the small probability of a two-photon component in the pulse. $V$ measures the degree of indistinguishability that the QD source delivers limited only by intrinsic phonon decoherence broadening of the zero-phonon line, which is the fundamental process limiting the performance. It is obtained according to \cite{santori2002,fox2018}
\begin{equation}
V = \frac{[1+2g^{(2)}(0)]\left(R^2+T^2\right) V_\mathrm{raw}}{2RT(1-\epsilon)^2}.
\end{equation}
where $R$ and $T$ are intensity reflection and transmission coefficients of the beam splitter, and $(1-\epsilon)$ is the classical visibility of the interferometer. 
This equation holds when the overall detection efficiency of the setup is much smaller than unity and when the two-photon contributions correspond to two fully distinguishable photons, which are good approximations in the present analysis. 
The effect of partially indistinguishable two-photon contributions and overall source, setup and detection efficiency approaching unity is to reduce the prefactor of 2 in front of $g^{(2)}(0)$ \cite{Eva2020}. 

\section{Evaluation of the efficiency of the single-photon source}
The overall efficiency of the single-photon source is determined by three parts: 1) the efficiency of the optical setup used to excite the QD and collect the single photons, 2) the QD source efficiency, which was introduced in the main text, and 3) the chip-to-fiber coupling efficiency.

\begin{figure*}
\centering
\includegraphics[width = \textwidth]{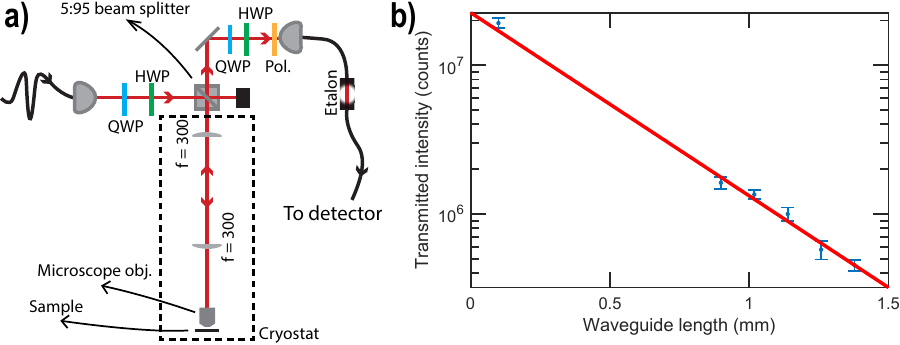}
\caption{(a)Schematic of the optical setup used in the excitation and collection of emission from a QD embedded in the nanophotonic device. The sample is cooled to a temperature of 1.6 K in a closed-cycle cryostat. A set of quarter (QWP) and half (HWP) wave plates are used to control the polarisation of the incident and collected light. (b) Transmitted intensity collected after propagation through waveguides of varying length, which is used to extract the propagation loss in the waveguide.}
\label{figS2}
\end{figure*}

\subsection{Setup efficiency}
The optical setup employed in our experiments is shown in Fig. \ref{figS2}(a).
The transmittance of each optical element used in the setup was carefully characterized using a continuous-wave narrow bandwidth diode laser operating at $947$ nm.
A resonant excitation laser is collimated and imaged to the back focal plane of a wide-field microscope objective (NA = 0.81; apochromat).
The microscope objective focuses the excitation laser to a diffraction-limited spot at the location of the QD in the sample.
The QD emission is collected at the right shallow-etched grating coupler (without the Y-splitter in Fig. \ref{figS1}(b)) using the same microscope objective.
The resonant laser and the collected emission is separated into different spatial modes using a 5:95 (reflection:transmission) beam splitter, where the transmission arm is used for collection.
The collected emission passes through a set of quarter and half wave plates (QWP, HWP in the figure) and is imaged onto a fibre collimator.
The linear polariser in the collection is aligned parallel to the polarisation axis of the collection grating outcoupler.
The collection efficiency of the imaging system $T$ from the device to the entrance of the collection fibre is $70\pm1\%$.
This efficiency is a product of the transmittances of the microscope objective (81\%), beam splitter (95\%), and other optical elements (5 each of on-average 98\%).
Additionally, we employ a spectral filter (linewidth = $3$ GHz) with a peak transmission efficiency $\eta_f$ = 87 $\pm$ 1\%, centered QD resonance to filter out the phonon side band.

\subsection{QD source efficiency}
In this section we evaluate the intrinsic efficiency of the QD source, which is determined by phonon decoherence, the single-photon coupling efficiency, and residual minor coupling to other exciton states.
We operate the QD at a gate voltage of 1.241 V, which ensures selective excitation of the neutral exciton $X_0$.
$X_0$ has two bright states corresponding to spectrally non-degenerate dipoles (fine structure splitting = 7.5 GHz) with orthogonal linear polarisations.
The location of the QD in the photonic crystal waveguide determines the coupling of the dipoles, quantified by the $\beta$-factor.
We measure an asymmetric coupling of the dipoles and extract the $\beta$-factor by fitting the resonant transmission dip using the following expression from Ref. \cite{javadi2015}
\begin{widetext}
\begin{equation}
T(\Delta\nu) = \frac{\{ (\Gamma + 2 \Gamma_d) [(\beta-1)^2 \Gamma + 2 \Gamma_d ] + 4 \Delta\nu^2\} (1 + \chi^2)} {(\Gamma + 2\Gamma_d)^2 + 4 \Delta \nu^2 + 4 \beta \Gamma \chi \Delta \nu  + \{ [ (\beta-1)\Gamma - 2 \Gamma_d]^2 + 4\Delta\nu^2 \} \chi^2},
\end{equation}
\end{widetext}
where, $\Delta\nu$ is the laser detuning from the QD resonance, $\Gamma$ is the natural linewidth of the QD, $\Gamma_d$ is the dephasing rate, and $\chi$ is the Fano parameter.
The Fano parameter accounts for the presence of weak back reflections at the waveguide terminations.
Of the governing parameters, we measure $\Gamma$ from time-resolved experiments (c.f. Fig. \ref{fig2}(c)). From the fit of the resonant transmission data, we find $\beta>90\%$ for the well-coupled dipole. This value is confirmed by the comparing the radiative lifetime of the QD in the waveguide to other QDs in a bulk part of the sample, from which we extract $\beta = (92 \pm 2)\%$ \cite{lodahl2014}.

\begin{figure*}
\centering
\includegraphics[width=\textwidth]{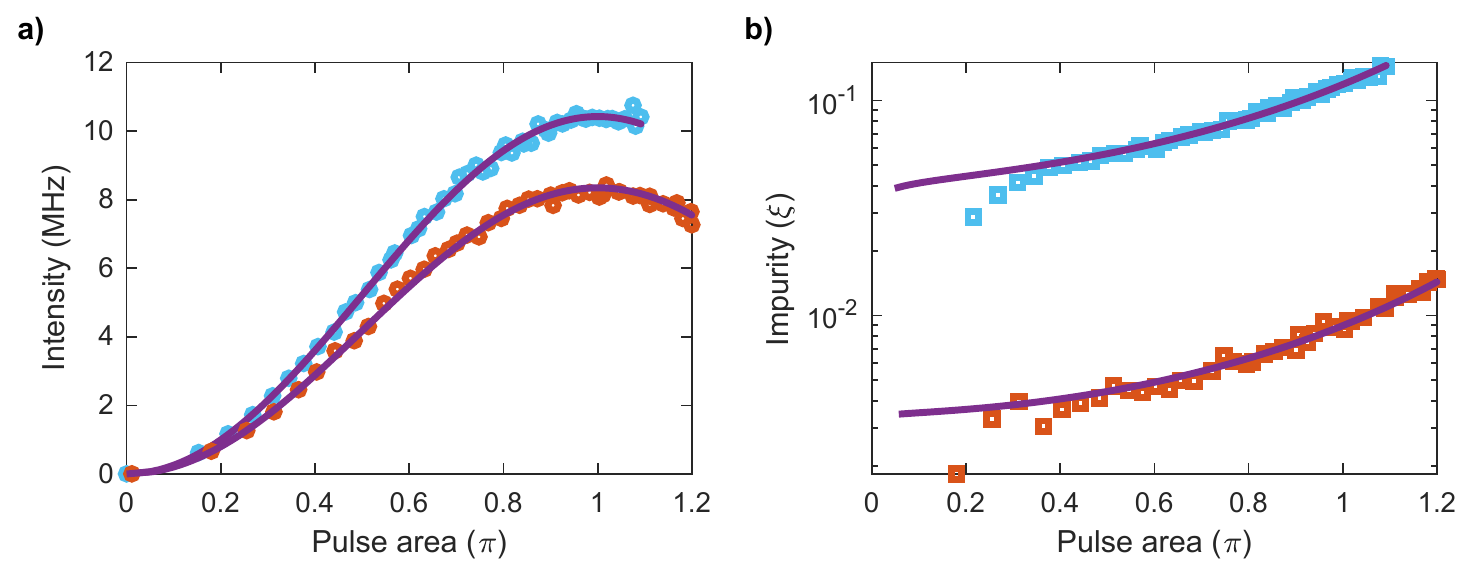}
\caption{(a) Pulsed resonance fluorescence from the QD with two different polarisations of the excitation laser: aligned to the well-coupled Y-dipole (blue markers) and aligned for the lowest single-photon impurity (red markers). (b) The measured single-photon impurity $\xi$ for the same polarisations as in (a).}
\label{figS3}
\end{figure*}

The well-coupled dipole can be selectively excited ($>$50:1 extinction of the pump laser) under pulsed resonant excitation by optimizing the polarisation of the pump laser that is focused on the QD.
However, the complex scattering of the excitation laser induced by the nanostructure can at times result in imperfect suppression of the excitation laser.
Residual laser scattering increases the single-photon impurity $\xi := I_\textrm{sp}/I_\textrm{laser}$, where $I_\textrm{sp}$ is the QD emission intensity and $I_\textrm{laser}$ is the residual laser intensity.
$\xi$ is related to the second-order correlation function as $g^{(2)}(0) = 2\xi - \xi^2$ \cite{kako2006}.
We first optimize the excitation polarisation for maximum $I_\textrm{sp}$ with the constraint that $\xi < 0.01$ at $\pi$-pulse operation.
For the QD studied here, this constraint implied that the polarisation of the excitation beam was chosen such that the well-coupled (Y) dipole was excited with a probability of $\eta_Y = 80 \%$ while the weaker coupled (X) dipole was excited with $\eta_X = 20 \%$ probability.
Under these conditions we observe a count rate of 8.3 MHz of single photons with $\xi = 0.007$ at $\pi$-pulse operation, cf. data in Fig. \ref{figS3}.
By changing the excitation polarisation we were able to increase the efficiency to observe a rate of $10.4$ MHz single photons with $\xi = 0.135$.
Another potential loss of population from the neutral exciton in the QD is coupling to non-radiative dark state.
This process will be revealed in time-dependent measurements of the second-order correlation function $g^{(2)}(\tau)$.
In the current experiment, this is a small effect and is quantified by modeling the very weak bunching observed in $g^{(2)}(\tau)$, cf. data in the inset of Fig. \ref{fig3}(d), using a 3-level system and calculating the dark state population \cite{Johansen2010,Davanco2014}.
We find that the resulting probability to decay on a radiative transition is  $\eta_\textrm{rad}$ = 98\%, i.e. $2\%$ blinking, which is likely a result of the dark exciton.

\begin{figure}
\centering
\includegraphics[width=\columnwidth]{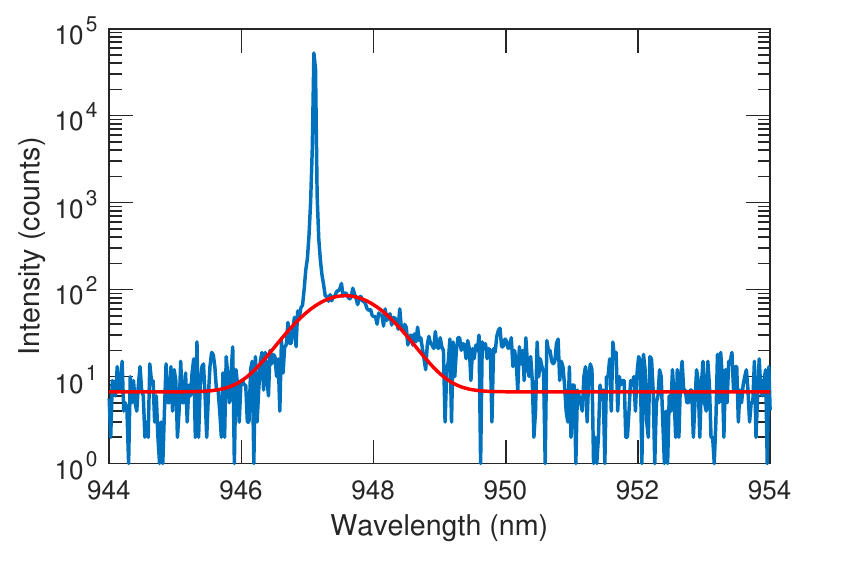}
\caption{Spectrally-resolved resonance fluorescence of the QD excited using a narrow bandwidth diode laser. The emission spectra are collected at an excitation power of $3$ nW and gate voltage of $1.24$V with the laser tuned to $\lambda$ = 947.4 nm. The red curve is the Gaussian fit of the emission in the phonon side band. The slight underestimation of the phonon sideband contribution (approximately $<1\%$) by the fit is due to the suppression of transmission at $\lambda >950$ nm due to the mode cutoff of the photonic crystal waveguide.}
\label{figS4}
\end{figure}

Finally, in order to generate highly indistinguishable photons, the phonon sidebands need to be filtered away spectrally.
The spectrally resolved emission with the excitation laser on resonance with the $Y$-dipole is shown in Fig. \ref{figS4}.
The resonance fluorescence spectrum exhibits a weak pedestal, which corresponds to the residual phonon side band.
The phonon sideband is fitted to a Gaussian to estimate the fraction emitted in the zero phonon line, $\eta_\textrm{zpl}$ = 95 $\pm$ 1\%.
The product of $\eta_Y$, $\beta$, $\eta_\textrm{zpl}$, and $\eta_\textrm{rad}$ is the intrinsic efficiency of the QD single photon source, which is found to reach up to 84 $\pm$ 4\%.
This corresponds to a single-photon rate of $122$ MHz for the operated repetition rate of the excitation laser of 145 MHz.
This is the key photonic resource provided by the device that is sufficient for reaching beyond the threshold of `quantum advantage' in a boson sampling experiment, as detailed in Sec. 5 below.

\subsection{Chip-to-fiber efficiency}
The photons emitted by the QD couple into the waveguide and propagate to the grating outcoupler, which diffracts them off-the chip in a narrow solid-angle (NA $\approx$ 0.2).
The propagation loss in single-mode nanobeam waveguides was estimated by measuring the transmission through waveguides of varying lengths.
Figure \ref{figS2}(b) shows the measured transmitted intensity at a fixed power for six waveguide lengths.
We fit the intensity decay to extract the propagation loss per unit length to be 10.5 dB/mm.
In order to estimate the possible additional propagation loss in the photonic crystal waveguide, we measure the ratio of the transmitted power in waveguides of equal length, with and without photonic crystal structure around the waveguide.
Using this measurement, we estimate the propagation loss in a photonic crystal waveguide to be 14 dB/mm.
Consequently, the propagation efficiency $\eta_p$ for the $10 \: \mu$m distance from the QD to the shallow-etch grating is $96\%$, which are the parameters for the device investigated in the present manuscript.

The diffraction efficiency of the shallow-etch grating couplers (SEG) was estimated following the method in Ref. \cite{zhou2018} to be $50\pm1\%$, which is slightly lower than the reported value in Ref. \cite{zhou2018}.
This reduction is caused by a small inaccuracy in the etch depth of the gratings and can be easily alleviated in a second run.
In order to efficiently collect the diffracted single photons into a single-mode fiber, optimal modematching $\eta_{g-f}$ of the diffracted mode to the fibre mode is necessary.
In our current implementation, we measure $\eta_{g-f} = 59\pm2\%$, which was limited due to the 4$f$-relay in the collection optics and can be readily improved to $>95\%$ with optimum choice of lenses.
The total efficiency of the generated single photons from QD to  the fiber is the product $\eta_p \eta_g \eta_{g-f}$, which in the current setup is $28\pm1\%$.
In the following, we present measurements on a next-generation SEG that enables $>90\%$ diffraction efficiency.

\begin{figure*}
\centering
\includegraphics[width=\textwidth]{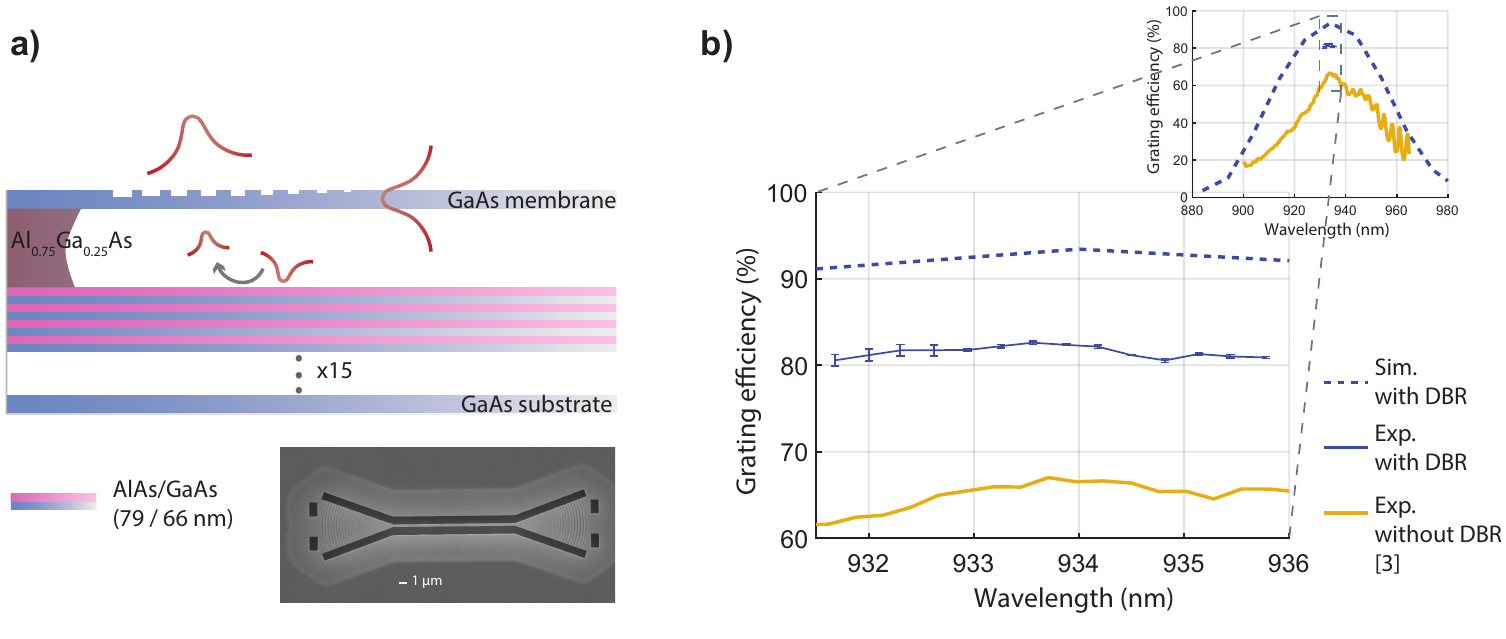}
\caption{(a) Schematics of the cross-section of a GaAs membrane (173 nm) with distributed Bragg reflectors AlAs/GaAs (79/66 nm) below the AlGaAs sacrificial layer (1150 nm). Inset: scanning electron microscope image of the nanostructure for characterization of the SEG coupling efficiency. (b) Measured grating efficiency (blue solid curve) for structures with DBR layer at 10 K, compared to simulated data (blue dashed curve). Measured efficiencies for structures without DBR are shown for comparison (yellow solid curve). Inset: full wavelength range of simulated data, indicating the grating transmission spectra. All lines are a guide for the eye, while the errorbars are extracted from the standard deviation over multiple structures. }
\label{figS5}
\end{figure*}

The coupling efficiency of the SEGs can be improved by the addition of a stack of distributed Bragg reflectors (DBR) below the AlGaAs sacrificial layer, as sketched in Fig. \ref{figS5}(a).
This extension boosts the reflection of the downwards-scattered light from 31 \% (bulk reflectivity of  GaAs) to $>99\%$.
To demonstrate this experimentally, a DBR stack comprising 15 layers of AlAs/GaAs (79/66 nm) was optimized using finite element calculations that resulted in an estimated coupling efficiency $>90 \%$ at the central wavelength of the SEG transmission spectra, as shown in the inset of Fig. \ref{figS5}(b) (blue dashed curve).
The wafers with the DBR layer were nanofabricated using a process similar to that described in Sec. I, but without the electrical contacts and characterized at 10 K in a liquid Helium flow cryostat.
The characterization structures, shown in Fig. \ref{figS5}(a), are made of two SEGs connected by a short waveguide (10 $\mu$m long).
The efficiency estimation is carried out using the same method as that employed in Ref. \cite{zhou2018}.
The SEG coupling efficiency is then obtained by normalizing the intensity of the light $I_\textrm{SEG}$ transmitted through two SEGs by the intensity $I_\textrm{Ref}$ of light directly reflected from the unpatterned, uniform surface in the following way:
\begin{equation}
\eta_\textrm{SEG}=\sqrt{\frac{I_\textrm{SEG}\cdot R_\textrm{DBR}}{I_\textrm{Ref}}},
\end{equation}
where $R_\textrm{DBR}$ is the reflection coefficient of the un-patterned surface with DBR, measured to be 95 \%, using a reflectometer measurement.
As shown in Fig. \ref{figS5}(b), the coupling efficiency around the central wavelength is as high as $\eta_\textrm{SEG}=$ 85 \% for one SEG (plain blue curve).
The error bars for each measured point arise from the statistical error, computed for measurements on three different structures.
The discrepancy between the measured and simulated values is due to the small beam size mismatch and asymmetry ($M^2 = 1.2$) in the SEG beam diameter, which can be compensated using beam circularizing optics.
Nonetheless, the measured efficiencies represent an improvement of $>20 \%$ compared to our previous work \cite{zhou2018}, and highlight the possibility of achieving chip-to-fiber coupling efficiencies $>90 \%$.

\subsection{Total efficiency}
\begin{table*}[ht]
\begin{center}
\begin{tabular}{|c|c|c|c|}
\hline
& Component  efficiency & Current device & Optimized value \\
\hline
\multirow{5}{*}{\rotatebox[origin=c]{90}{SOURCE}} & $\beta$ & $92 \pm 5\%$ & $>99 \%$ \cite{lodahl2014,Scarpelli2019}\\
&$\eta_Y$ & $>98\%$ & $100\%$ \\
&Zero phonon line $\eta_\textrm{zpl}$ & $95 \pm 1 \%$ & $95\%$ \cite{sorensen2018}\\
&Radiative $\eta_\textrm{rad}$ & $98\%$ & $>98\%$ \\
\cline{2-4}
& Single-photon source efficiency $\eta_S$  & $84\%$ & $>92\%$\\
\hline
\hline
\multirow{5}{*}{\rotatebox[origin=c]{90}{SETUP}} &Directionality & $50\%$ & $100 \%$\cite{lodahl2015}\\
&Collection optics $T$ & $70 \pm 1\%$ & $100 \%$ \\
&Spectral filter $\eta_f$ & $87 \pm 1\%$ & $>98\%$ \\
&On-chip propagation $\eta_p$ & 96\% & 96\% \\
&Chip-to-fiber $\eta_g\eta_{g-f}$ & $29 \pm 1\%$ & $>90\%$ \\
\hline
&Expected single-photon rate & $10.3 \pm 0.7$ MHz &  $>114$ MHz\\
&Measured single-photon rate & $10.4$ MHz & \\
\hline
\end{tabular}
\end{center}
\caption{Breakdown of the efficiencies of the source and the characterization setup.}
\end{table*}

Table S1 summarizes the efficiencies of the device and the characterization setup that were discussed above.
From all the measured parameters we expect a single-photon rate of the fiber-coupled source of $(10.3 \pm 0.7)$ MHz, which matches very well with the measured rate of 10.4 MHz.
Note that the characterized device was two-ended meaning that only half of the generated photons were collected (Directionality: $50 \%$).
All efficiencies are therefore fully accounted for in the experiment, and the second column of Table S1 breaks down the parameters of a fully-optimized system using values already experimentally achieved or readily projected.
It is found that a fiber-coupled source of indistinguishable photons exceeding an overall efficiency $78 \%$ can be achieved, which exceeds the requirements of demonstrating quantum advantage.
It could be mentioned that the chip-to-fiber efficiency could be improved even further, e.g., by replacing the grating outcouplers with tapered-waveguide coupling into tapered optical fibers, where efficiencies exceeding $96\%$ have been achieved \cite{Pu2010,Tiecke2015}.

\section{Scalable boson sampling with a QD single-photon source}
Boson sampling has been recognized as one of the most promising ways to establish the advantage of a quantum machine over its classical counterpart \cite{Aaronson2013}.
Photonic boson sampling consists of $N$ identical photons that undergo linear optical transformation in an optical network with $M$ optical modes (mathematically, an $M\times M$ unitary matrix $U$) and are subsequently detected.
The task of sampling the output distribution of the network is efficiently performed in a quantum machine implementing the sampling rather than simulating the experiment on a classical computer.
The classical complexity of this task can be understood as follows:
Let $\vec{k}$ and $\vec{l}$, both $M$-dimensional vectors, denote the photon occupation number at each optical mode, such that $\sum_{m=1}^{M} k_m = N$ and $\sum_{m=1}^{M} l_m = N$.
The probability of detecting the output configuration $\vec{l}$, given the input configuration $\vec{k}$ in the case of perfectly indistinguishable photons is \cite{Aaronson2013}
\begin{equation}
\hat{p}^{(id)} (\vec{l}|\vec{k}) = \frac{| \textrm{perm}( U[\vec{k}|\vec{l}])|^2}{\mu(\vec{k}) \mu(\vec{l})},
\end{equation}
where, $U[\vec{k}|\vec{l}]$ is the sub-matrix of $U$ with rows specified from $\vec{k}$ and columns from $\vec{l}$ and $\mu(\vec{k}) := \prod_{m=1}^M k_m!$.
The calculation of permanents on a classical computer is conjectured to be hard with runtime scaling exponentially with the size of the sub-matrix.
In contrast, the corresponding probability for distinguishable photons at the input is \cite{Aaronson2013}
\begin{equation}
\hat{p}^{(dist)} (\vec{l}|\vec{k}) = \frac{\textrm{perm}(|U[\vec{k}|\vec{l}]|^2)}{\mu(\vec{k})}.
\end{equation}
The permanent of the above absolute-squared-matrix, which has non-negative matrix elements, can be efficiently approximated (polynomial-time with increasing matrix size) on a classical computer \cite{jerrum2004}.

Experimental realizations of boson sampling can only be operated in a near-ideal regime.
Several theoretical studies have investigated the effect of realistic imperfections in the network \cite{Arkhipov2015} as well as the internal state of the photons, i.e. non-identical photons \cite{Shchesnovich2014,Tichy2015,Rhode2015,Shchesnovich2015} and photon loss\cite{Aaronson2016,Wang2018}.
These studies proposed approximate classical algorithms for boson sampling that could scale efficiently depending on the degree of photon distinguishability and photon loss \cite{Keshari2016,Neville2017,Clifford2018,Renema2018,Oszmaniec_2018,Patron2019,Moylett2020}.
We employ these studies to investigate the applicability of our source in the quantum advantage regime.

\begin{figure}
\centering
\includegraphics[width=\columnwidth]{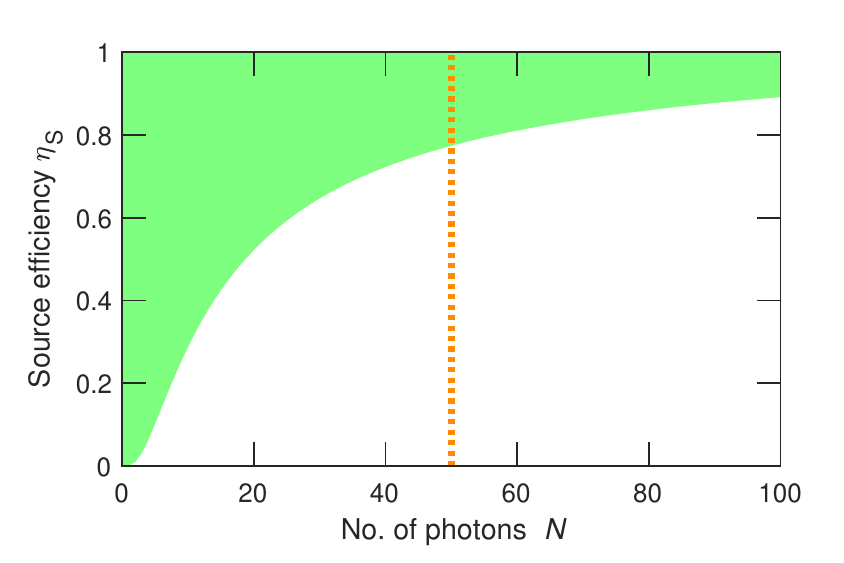}
\caption{Source efficiency $\eta_S$ required for $N$-photon interference.
The green area is the regime where classical algorithms cannot approximate the $N$ photon interference with $k < N$ photons for a photon indistinguihsability $V = 0.96$.
We set an error-tolerance $E \leq 0.001$ for the approximation algorithm.
The dotted line represents the expected quantum advantage threshold, $N\sim50$ photons.}
\label{figS7}
\end{figure}

The single-photon pulse train generated from a QD can be demultiplexed into different spatial modes \cite{pan2017,pan2019boson,lodahl2019demux} for use in a boson sampling experiment.
The two imperfections in single photon sources that affect scalability of boson sampling are non-unity pairwise photon indistinguishability and efficiency.
The pairwise photon indistinguishability is represented as the matrix $\mathcal{S}$ with elements $S_{ij} = \langle \phi_i | \phi_j \rangle \forall \{i,j\} \in \{1, \cdots, N\}$, where $\phi_i$ and $\phi_j$ are the internal states of the $i$-th and $j$-th photons at the input, respectively.
For perfectly indistinguishable photons, $S_{i,j} = 1$.
For real sources, $|S_{i,j}|^2$ is the measured HOM visibility $V_{ij}$ between photons $i$ and $j$ in the pulse train.
For a QD single-photon source, $S_{i,j} = \sqrt{V_{ij}}$ as the photons in the pulse train are spontaneously emitted and hence lack phase coherence between each other.
The source efficiency $\eta_S$ is uniform for all the photons in the single-photon pulse train.

Efficient classical algorithms approximate boson sampling with partially distinguishable and lossy photon sources.
These algorithms approximate $N$-photon interference of imperfect photons as interference between $k$ perfect single photons ($k < N$) within a certain approximation error $E$.
$k$ is thus the approximation order of the algorithm, with $k = N$ indicating no approximation.
The bound on the error $E$  at an approximation order $k$ is related to the source imperfections as \cite{Renema2018}
\begin{equation}
 E < \sqrt{\frac {(\eta_S V)^{k+1}}{1-\eta_S V}},
\label{eq:eqBS}
\end{equation}
where, $V = \textrm{max.}(V_{ij})$ and $k$ is the approximation order i.e. the reduced number of photons interfering in the network ($0\leq k \leq N$).
Further, we assume unity transmission efficiency of the network $\eta_{net} = 1$ and detection $\eta_{det} = 1$.
Therefore, for a given $E$, $V$ and $\eta_S$, the inequality in the above equation sets the upper limit on the number of photons that can be classically simulated.
The interface between the green and the white shaded areas in Fig. \ref{figS7} is the minimum $\eta_S$ required such that $k = N$ for a single-photon source with $V = 0.96$ and a classical error $E = 0.001$.
The quantum advantage regime has been identified as the photonic resource necessary to outperform the classical computation of the occurence probability $\hat{p}(\vec{l}|\vec{k})$ on a supercomputer.
This regime is expected to be $N\sim50$ (marked with the dashed line) \cite{Neville2017}.

\begin{figure*}
\centering
\includegraphics[width=\textwidth]{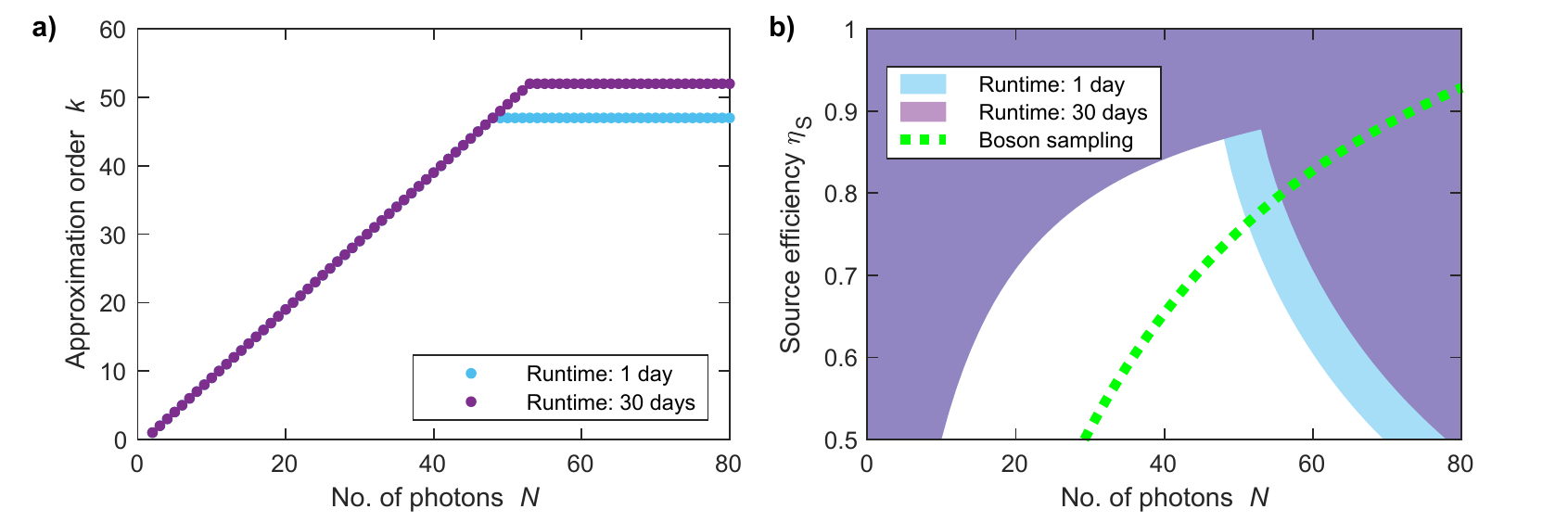}
\caption{(a) Maximum approximation order $k$ for a classical algorithm that can be performed in a limited runtime of 1 or 30 days on a supercomputer with 100 PFLOPS sustained operation.
(b) Source efficiency $\eta_S$ that can be classically approximated with $k \leq N-1$ for a source with $V = 0.96$ within a fixed runtime is marked by the white region.
The shaded region is $\eta_S$ to achieve exponential scaling of computational hardness in boson sampling, i.e. $R_q/R_c >1$.
We set the classical error-tolerance $E = 0.001$.
The dashed curve is the minimum $\eta_S$ required to experimentally perform the boson sampling within a runtime of 30 days.
Technologically, the quantum advantage threshold is the minimum $\eta_S$ where the boson sampler can outperform the classical algorithm.
At a fixed runtime of 30 days, this occurs at the intersection of the purple shaded region with the dashed curve at $N = 54$ photons for $\eta_S = 0.78$.}
\label{figS8}
\end{figure*}

The performance comparison is quantified from the runtimes of the boson sampler $R_q$ and the classical algorithm $R_c$ such that $R_q/R_c>1$ signifies quantum advantage.
The upper-bound on $R_q$ is currently set by the limit on the continuous operation of the photon sources, which in practice is weeks \cite{pan2019boson,Uppu2020}.
Given this finite runtime, we calculate $R_c$ required to calculate $k$-dimensional matrix permanents on a classical super computer operating at a sustained rate of 100 PFLOPS \cite{Summit2019}.
We set the number of permanents to be calculated as the number of multiphoton detection events $N_s$ that would be accumulated from the boson sampler to reject the distinguishable sampler hypothesis.
A practical estimate of $N_s$ is given by the coupon collector's problem for distinguishable photons, which is the number of multiphoton detection events required to sample at least one photon from each of the $M$ output modes, $N_s \approx M \log(M)/N$ \cite{Ferrante2012}.

Given the upper-bound on $R_c$, we calculate the highest approximation order $k$ for $N$ photons that can be realized for a classical algorithm based on state truncation\cite{Moylett2020}, which is shown in Fig. \ref{figS8}(a) for runtimes of 1 day and 30 days.
A linear increase with $k = N-1$ is observed up to $N = 52$ (close to the previously expected quantum advantage bound), beyond which $k$ is constant.
This limit on $k$ is due to the exponential increase in the computation time that scales with $k$ as $\approx k2^k$.
If we fix the photon indistinguishability $V$ and set a classical error-tolerance $E \leq 0.001$, we can extract the highest efficiency $\eta_S$ that can be classically computed within the limited $k$ in the finite runtime \cite{Moylett2020}.
This value of $\eta_S$ at every $N$ should be compared with the $\eta_S$ required to detect $N_s$ multiphoton events from a boson sampler within the runtime to determine the regime of quantum advantage.

The efficiency of a boson sampler is a product of four component efficiencies: 1) source efficiency $\eta_S$, 2) demultiplexing efficiency $\eta_{dx}$, 3) optical network transmission $\eta_{net}$, and detection efficiency $\eta_{d}$.
In a realistic assessment of the experimental feasibility, we determine $\eta_S$ required for quantum advantage and fix the other efficiencies to experimentally realizable values of $\eta_{dx} = 90\%$, $\eta_{net} = 92\%$, and $\eta_d = 92\%$ \cite{pan2019boson,zwiller2017}.
The white regions in Fig. \ref{figS8}(b) demarcate $\eta_S$ that can be classically simulated for a source with $V = 0.96$.
The upper-bound on classically-simulatable $\eta_S$ for the runtime of 30 days closely follows the estimate in Fig. \ref{figS7} up to $N = 52$, after which it decreases monotonically due to the increasing $N-k$.
To estimate the lower-bound on $\eta_S$ for an experimental boson sampler to collect $N_s$ collision-free multiphoton events, we use the probability of their occurrence over the Haar measure \cite{Arkhipov2012}
\begin{equation}
P^{(id)} \equiv \sum_{\vec{l}|l_i\in \{0,1\}} \hat{p}^{(id)}(\vec{l}|\vec{k}) = \binom{M}{N}/\binom{M+N-1}{N}.
\end{equation}
The runtime of the boson sampler is then calculated by assuming a $1$ GHz repetition rate of the laser exciting the QD single-photon source, i.e. the multiphoton generation rate using active demultiplexing is $1/N$ GHz.
The dashed curve in Fig. \ref{figS8}(b) (also plotted in Fig. \ref{fig3}(d)) is $\eta_S$ required for collecting $N_s$ multiphoton events from a boson sampler, which monotonically increases with $N$.
Comparing the thresholds on $\eta_S$ set by the classical algorithm (runtime of 30 days) and the boson sampler, we conclude that the minimum $\eta_S$ for quantum advantange is $78\%$ at $N = 54$ photons.

\subsection{Effect of source distinguishability on scalability}
\begin{figure*}[ht]
\includegraphics[width=\textwidth]{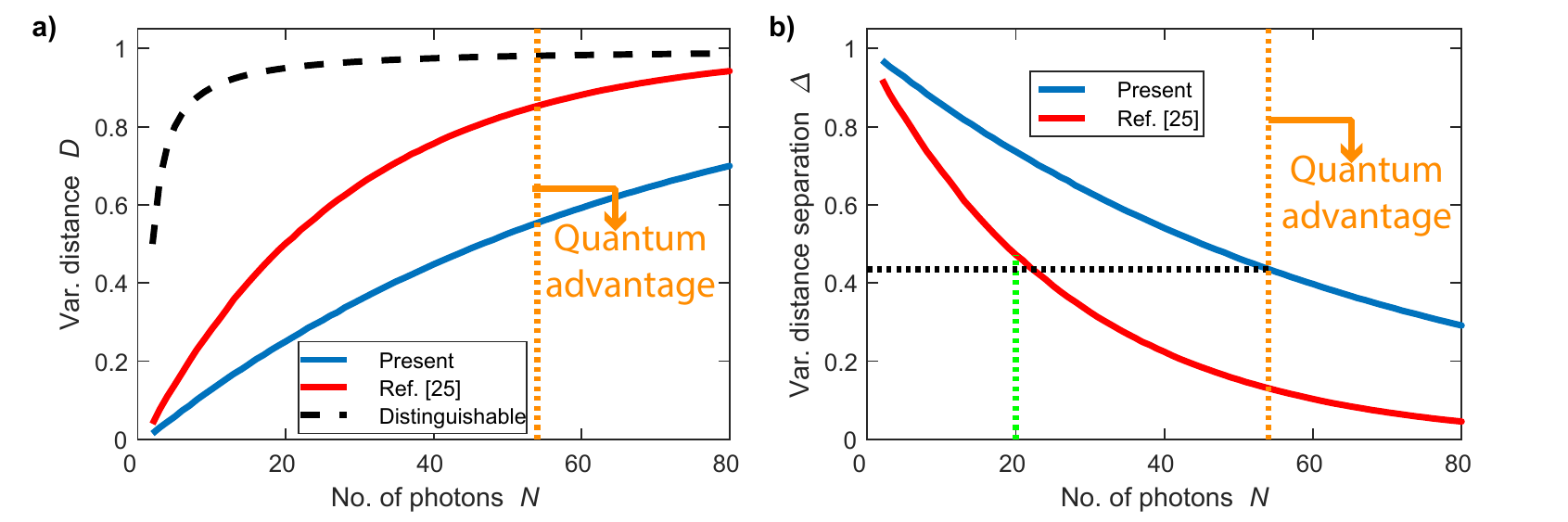}
\caption{(a) Variational distances for two real sources (blue curve - present source; red curve - source employed in Ref. \cite{pan2019boson}) from perfectly indistinguishable photon source when employed for boson sampling. The upper bound on the variational distance from an ideal boson sampler is set by the case with distinguishable photons (dashed curve). (b) Variational distance separation $\Delta$ for the two sources in (a) highlights the possibility to perform boson sampling at the quantum advantage threshold ($N=54$ photons) using our source at a comparable $\Delta$ as that used for $N=20$ photons in Ref. \cite{pan2019boson}.}
\label{figS9}
\end{figure*}

The reliability of boson sampling validation techniques \cite{Spagnolo2014,Carolan2014,Liu2016,Agresti2019} at a fixed $N_s$ relies on the indistinguishability of the photons.
We will use the validation against the case of distinguishable photons as an example in this section.
These validation schemes quantify the dissimilarity of the $N$-photon output probability distribution in the network $\mathbf{p} = \{\hat{p}(\vec{l}|\vec{k})\}$ from that of distinguishable photons $\mathbf{p^{(dist)}} = \{\hat{p}^{(dist)}(\vec{l}|\vec{k})\}$.
The estimate of $N_s$ employed earlier is sufficient for validating a boson sampler injected with indistinguishable photons, i.e. the maximally dissimilar distribution $\mathbf{p^{(id)}} = \{\hat{p}^{(id)}(\vec{l}|\vec{k})\}$.
With partial photon distinguishability, the distibutions $\mathbf{p}$ and $\mathbf{p^{(id)}}$ are ``distant'' is probability space.
Given the photon distinguishability matrix $\mathcal{S}$, the closeness of $\mathbf{p}$ and $\mathbf{p^{(id)}}$ is quantified using the variational distance, which is defined as
\begin{equation}
 D(\mathbf{p}^{(id)}, \mathbf{p}) = \frac{1}{2} \sum_{\vec{l}} |\hat{p}^{(id)}(\vec{l}|\vec{k}) - \hat{p}(\vec{l}|\vec{k})|.
\end{equation}
The upper bound (conjectured to be a tight bound) for the variational distance with photon distinguishability has been derived to be \cite{Shchesnovich2015}
\begin{equation}
 D(\mathbf{p}^{(id)}, \mathbf{p}) \leq 1 - \frac{\textrm{ perm}(\mathcal{S})}{N!}.
\end{equation}

Recent boson sampling experiments with up to $20$ input photons \cite{pan2019boson} have employed a source where $V$ decreased over the single-photon pulse train that results in a non-uniform pairwise visibility $V_{ij}$.
In contrast, the improved source demonstrated here exhibits $V_{ij} = V ~ \forall i \neq j$.
This improvement in $V$ directly impacts the scaling of the variational distance with $N$ as seen in Fig. \ref{figS9}(a).
For the validation of boson sampling, the Bayesian prior employed is based on the extreme cases of perfectly indistinguishable and distinguishable photons.
The variational distance between these extreme cases (dashed curve) for Haar unitary matrices is $D(\mathbf{p}^{(id)}, \mathbf{p}^{(dist)}) \approx (N-1)/N$ \cite{Tichy2015}.
Non-unity variational distance separation $\Delta \equiv 1 - D(\mathbf{p}^{(id)},\mathbf{p})/D(\mathbf{p}^{(id)}, \mathbf{p}^{(dist)})$ decreases the reliability of validation method for a fixed $N_s$.
In Ref. \cite{pan2019boson},  $N=20$ photon (14 detected photons) boson sampling in a 60-mode network was reported to achieve $99.9\%$ reliability with $\approx$30 multiphoton events.
In comparison, the coupon collector's problems estimate of the number of events $N_s \approx 15$, which indicates a $2\times$ overhead.
The non-unity $\Delta = 0.47$ for the source employed for boson sampling results in this overhead.
If the same source was employed at the quantum advantage threshold $N = 54$ photons, we observe that $\Delta = 0.13$ causing a further increase in the overhead.
Therefore, the validation of boson sampling could require much higher number of events than those used to establish the quantum advantage threshold in Fig. \ref{figS8}(b).
The collection of these additional events require longer runtimes and thus inhibits the scaling up into the quantum advantage regime.
Using our improved source, boson sampling with $N = 54$ photons has a $\Delta = 0.44$, cf. Fig. \ref{figS9}(b).
This is a comparable $\Delta$ to the one reported in Ref. \cite{pan2019boson} with $N=20$.
Consequently, the improved performance of our source means that validation can be implemented at the quantum advantage threshold with an overhead similar to the previous experiments that can be readily achieved using the demonstrated $\eta_S$, cf. Appendix D.4.

\begin{acknowledgments}
We thank Prof. Richard Warbuton (University of Basel) for discussions and help on designing low-noise hetrostructures.

The authors gratefully acknowledge financial support from Danmarks Grundforskningsfond (DNRF) (Center for Hybrid Quantum Networks (Hy-Q; DNRF139)), H2020 European Research Council (ERC) (SCALE), Styrelsen for Forskning og Innovation (FI) (5072-00016B QUANTECH), Bundesministerium f\"{u}r Bildung und Forschung (BMBF) (16KIS0867, Q.Link.X), Deutsche Forschungsgemeinschaft (DFG) (TRR 160).
\end{acknowledgments}


\begin{thebibliography}{71}%
\makeatletter
\providecommand \@ifxundefined [1]{%
 \@ifx{#1\undefined}
}%
\providecommand \@ifnum [1]{%
 \ifnum #1\expandafter \@firstoftwo
 \else \expandafter \@secondoftwo
 \fi
}%
\providecommand \@ifx [1]{%
 \ifx #1\expandafter \@firstoftwo
 \else \expandafter \@secondoftwo
 \fi
}%
\providecommand \natexlab [1]{#1}%
\providecommand \enquote  [1]{``#1''}%
\providecommand \bibnamefont  [1]{#1}%
\providecommand \bibfnamefont [1]{#1}%
\providecommand \citenamefont [1]{#1}%
\providecommand \href@noop [0]{\@secondoftwo}%
\providecommand \href [0]{\begingroup \@sanitize@url \@href}%
\providecommand \@href[1]{\@@startlink{#1}\@@href}%
\providecommand \@@href[1]{\endgroup#1\@@endlink}%
\providecommand \@sanitize@url [0]{\catcode `\\12\catcode `\$12\catcode
  `\&12\catcode `\#12\catcode `\^12\catcode `\_12\catcode `\%12\relax}%
\providecommand \@@startlink[1]{}%
\providecommand \@@endlink[0]{}%
\providecommand \url  [0]{\begingroup\@sanitize@url \@url }%
\providecommand \@url [1]{\endgroup\@href {#1}{\urlprefix }}%
\providecommand \urlprefix  [0]{URL }%
\providecommand \Eprint [0]{\href }%
\providecommand \doibase [0]{http://dx.doi.org/}%
\providecommand \selectlanguage [0]{\@gobble}%
\providecommand \bibinfo  [0]{\@secondoftwo}%
\providecommand \bibfield  [0]{\@secondoftwo}%
\providecommand \translation [1]{[#1]}%
\providecommand \BibitemOpen [0]{}%
\providecommand \bibitemStop [0]{}%
\providecommand \bibitemNoStop [0]{.\EOS\space}%
\providecommand \EOS [0]{\spacefactor3000\relax}%
\providecommand \BibitemShut  [1]{\csname bibitem#1\endcsname}%
\let\auto@bib@innerbib\@empty
\bibitem [{\citenamefont {Aspuru-Guzik}\ and\ \citenamefont
  {Walther}(2012)}]{walther2012}%
  \BibitemOpen
  \bibfield  {author} {\bibinfo {author} {\bibfnamefont {A.}~\bibnamefont
  {Aspuru-Guzik}}\ and\ \bibinfo {author} {\bibfnamefont {P.}~\bibnamefont
  {Walther}},\ }\href@noop {} {\bibfield  {journal} {\bibinfo  {journal} {Nat.
  Phys.}\ }\textbf {\bibinfo {volume} {8}},\ \bibinfo {pages} {285} (\bibinfo
  {year} {2012})}\BibitemShut {NoStop}%
\bibitem [{\citenamefont {M{\'a}ttar}\ \emph {et~al.}(2018)\citenamefont
  {M{\'a}ttar}, \citenamefont {Ko{\l}ody{\'n}ski}, \citenamefont {Skrzypczyk},
  \citenamefont {Cavalcanti}, \citenamefont {Banaszek},\ and\ \citenamefont
  {Ac{\'\i}n}}]{acin2018}%
  \BibitemOpen
  \bibfield  {author} {\bibinfo {author} {\bibfnamefont {A.}~\bibnamefont
  {M{\'a}ttar}}, \bibinfo {author} {\bibfnamefont {J.}~\bibnamefont
  {Ko{\l}ody{\'n}ski}}, \bibinfo {author} {\bibfnamefont {P.}~\bibnamefont
  {Skrzypczyk}}, \bibinfo {author} {\bibfnamefont {D.}~\bibnamefont
  {Cavalcanti}}, \bibinfo {author} {\bibfnamefont {K.}~\bibnamefont
  {Banaszek}}, \ and\ \bibinfo {author} {\bibfnamefont {A.}~\bibnamefont
  {Ac{\'\i}n}},\ }\href@noop {} {\bibfield  {journal} {\bibinfo  {journal}
  {arXiv:1803.07089}\ } (\bibinfo {year} {2018})}\BibitemShut {NoStop}%
\bibitem [{\citenamefont {Borregaard}\ \emph {et~al.}(2019)\citenamefont
  {Borregaard}, \citenamefont {Pichler}, \citenamefont {Sch{\"o}der},
  \citenamefont {Lukin}, \citenamefont {Lodahl},\ and\ \citenamefont
  {S{\o}rensen}}]{Borregaard2019}%
  \BibitemOpen
  \bibfield  {author} {\bibinfo {author} {\bibfnamefont {J.}~\bibnamefont
  {Borregaard}}, \bibinfo {author} {\bibfnamefont {H.}~\bibnamefont {Pichler}},
  \bibinfo {author} {\bibfnamefont {T.}~\bibnamefont {Sch{\"o}der}}, \bibinfo
  {author} {\bibfnamefont {M.~D.}\ \bibnamefont {Lukin}}, \bibinfo {author}
  {\bibfnamefont {P.}~\bibnamefont {Lodahl}}, \ and\ \bibinfo {author}
  {\bibfnamefont {A.~S.}\ \bibnamefont {S{\o}rensen}},\ }\href@noop {}
  {\bibfield  {journal} {\bibinfo  {journal} {arXiv:1907.05101}\ } (\bibinfo
  {year} {2019})}\BibitemShut {NoStop}%
\bibitem [{\citenamefont {Lindner}\ and\ \citenamefont
  {Rudolph}(2009)}]{Lindner2009}%
  \BibitemOpen
  \bibfield  {author} {\bibinfo {author} {\bibfnamefont {N.~H.}\ \bibnamefont
  {Lindner}}\ and\ \bibinfo {author} {\bibfnamefont {T.}~\bibnamefont
  {Rudolph}},\ }\href@noop {} {\bibfield  {journal} {\bibinfo  {journal} {Phys.
  Rev. Lett.}\ }\textbf {\bibinfo {volume} {103}},\ \bibinfo {pages} {113602}
  (\bibinfo {year} {2009})}\BibitemShut {NoStop}%
\bibitem [{\citenamefont {Kimble}(2008)}]{kimble2008}%
  \BibitemOpen
  \bibfield  {author} {\bibinfo {author} {\bibfnamefont {H.~J.}\ \bibnamefont
  {Kimble}},\ }\href@noop {} {\bibfield  {journal} {\bibinfo  {journal}
  {Nature}\ }\textbf {\bibinfo {volume} {453}},\ \bibinfo {pages} {1023}
  (\bibinfo {year} {2008})}\BibitemShut {NoStop}%
\bibitem [{\citenamefont {Wehner}\ \emph {et~al.}(2018)\citenamefont {Wehner},
  \citenamefont {Elkouss},\ and\ \citenamefont {Hanson}}]{hanson2018}%
  \BibitemOpen
  \bibfield  {author} {\bibinfo {author} {\bibfnamefont {S.}~\bibnamefont
  {Wehner}}, \bibinfo {author} {\bibfnamefont {D.}~\bibnamefont {Elkouss}}, \
  and\ \bibinfo {author} {\bibfnamefont {R.}~\bibnamefont {Hanson}},\
  }\href@noop {} {\bibfield  {journal} {\bibinfo  {journal} {Science}\ }\textbf
  {\bibinfo {volume} {362}},\ \bibinfo {pages} {eaam9288} (\bibinfo {year}
  {2018})}\BibitemShut {NoStop}%
\bibitem [{\citenamefont {Rudolph}(2017)}]{Rudolph2017}%
  \BibitemOpen
  \bibfield  {author} {\bibinfo {author} {\bibfnamefont {T.}~\bibnamefont
  {Rudolph}},\ }\href@noop {} {\bibfield  {journal} {\bibinfo  {journal} {APL
  Photonics}\ }\textbf {\bibinfo {volume} {2}},\ \bibinfo {pages} {030901}
  (\bibinfo {year} {2017})}\BibitemShut {NoStop}%
\bibitem [{\citenamefont {Monroe}\ and\ \citenamefont
  {Kim}(2013)}]{monroe2013}%
  \BibitemOpen
  \bibfield  {author} {\bibinfo {author} {\bibfnamefont {C.}~\bibnamefont
  {Monroe}}\ and\ \bibinfo {author} {\bibfnamefont {J.}~\bibnamefont {Kim}},\
  }\href@noop {} {\bibfield  {journal} {\bibinfo  {journal} {Science}\ }\textbf
  {\bibinfo {volume} {339}},\ \bibinfo {pages} {1164} (\bibinfo {year}
  {2013})}\BibitemShut {NoStop}%
\bibitem [{\citenamefont {Kaneda}\ and\ \citenamefont
  {Kwiat}(2019)}]{kwiat2019}%
  \BibitemOpen
  \bibfield  {author} {\bibinfo {author} {\bibfnamefont {F.}~\bibnamefont
  {Kaneda}}\ and\ \bibinfo {author} {\bibfnamefont {P.~G.}\ \bibnamefont
  {Kwiat}},\ }\href@noop {} {\bibfield  {journal} {\bibinfo  {journal} {Sci.
  Adv.}\ }\textbf {\bibinfo {volume} {5}},\ \bibinfo {pages} {eaaw8586}
  (\bibinfo {year} {2019})}\BibitemShut {NoStop}%
\bibitem [{\citenamefont {Arcari}\ \emph {et~al.}(2014)\citenamefont {Arcari},
  \citenamefont {S\"ollner}, \citenamefont {Javadi}, \citenamefont
  {Lindskov~Hansen}, \citenamefont {Mahmoodian}, \citenamefont {Liu},
  \citenamefont {Thyrrestrup}, \citenamefont {Lee}, \citenamefont {Song},
  \citenamefont {Stobbe},\ and\ \citenamefont {Lodahl}}]{lodahl2014}%
  \BibitemOpen
  \bibfield  {author} {\bibinfo {author} {\bibfnamefont {M.}~\bibnamefont
  {Arcari}}, \bibinfo {author} {\bibfnamefont {I.}~\bibnamefont {S\"ollner}},
  \bibinfo {author} {\bibfnamefont {A.}~\bibnamefont {Javadi}}, \bibinfo
  {author} {\bibfnamefont {S.}~\bibnamefont {Lindskov~Hansen}}, \bibinfo
  {author} {\bibfnamefont {S.}~\bibnamefont {Mahmoodian}}, \bibinfo {author}
  {\bibfnamefont {J.}~\bibnamefont {Liu}}, \bibinfo {author} {\bibfnamefont
  {H.}~\bibnamefont {Thyrrestrup}}, \bibinfo {author} {\bibfnamefont {E.~H.}\
  \bibnamefont {Lee}}, \bibinfo {author} {\bibfnamefont {J.~D.}\ \bibnamefont
  {Song}}, \bibinfo {author} {\bibfnamefont {S.}~\bibnamefont {Stobbe}}, \ and\
  \bibinfo {author} {\bibfnamefont {P.}~\bibnamefont {Lodahl}},\ }\href@noop {}
  {\bibfield  {journal} {\bibinfo  {journal} {Phys. Rev. Lett.}\ }\textbf
  {\bibinfo {volume} {113}},\ \bibinfo {pages} {093603} (\bibinfo {year}
  {2014})}\BibitemShut {NoStop}%
\bibitem [{\citenamefont {Lodahl}\ \emph {et~al.}(2015)\citenamefont {Lodahl},
  \citenamefont {Mahmoodian},\ and\ \citenamefont {Stobbe}}]{lodahl2015}%
  \BibitemOpen
  \bibfield  {author} {\bibinfo {author} {\bibfnamefont {P.}~\bibnamefont
  {Lodahl}}, \bibinfo {author} {\bibfnamefont {S.}~\bibnamefont {Mahmoodian}},
  \ and\ \bibinfo {author} {\bibfnamefont {S.}~\bibnamefont {Stobbe}},\
  }\href@noop {} {\bibfield  {journal} {\bibinfo  {journal} {Rev. Mod. Phys.}\
  }\textbf {\bibinfo {volume} {87}},\ \bibinfo {pages} {347} (\bibinfo {year}
  {2015})}\BibitemShut {NoStop}%
\bibitem [{\citenamefont {Somaschi}\ \emph {et~al.}(2016)\citenamefont
  {Somaschi}, \citenamefont {Giesz}, \citenamefont {De~Santis}, \citenamefont
  {Loredo}, \citenamefont {Almeida}, \citenamefont {Hornecker}, \citenamefont
  {Portalupi}, \citenamefont {Grange}, \citenamefont {Ant{\'o}n}, \citenamefont
  {Demory}, \citenamefont {G\'{o}mez}, \citenamefont {I.}, \citenamefont
  {Lanzillotti-Kimura}, \citenamefont {Lema\'{i}tre}, \citenamefont {A.},
  \citenamefont {White}, \citenamefont {Lanco},\ and\ \citenamefont
  {Senellart}}]{senellart2016}%
  \BibitemOpen
  \bibfield  {author} {\bibinfo {author} {\bibfnamefont {N.}~\bibnamefont
  {Somaschi}}, \bibinfo {author} {\bibfnamefont {V.}~\bibnamefont {Giesz}},
  \bibinfo {author} {\bibfnamefont {L.}~\bibnamefont {De~Santis}}, \bibinfo
  {author} {\bibfnamefont {J.}~\bibnamefont {Loredo}}, \bibinfo {author}
  {\bibfnamefont {M.~P.}\ \bibnamefont {Almeida}}, \bibinfo {author}
  {\bibfnamefont {G.}~\bibnamefont {Hornecker}}, \bibinfo {author}
  {\bibfnamefont {S.~L.}\ \bibnamefont {Portalupi}}, \bibinfo {author}
  {\bibfnamefont {T.}~\bibnamefont {Grange}}, \bibinfo {author} {\bibfnamefont
  {C.}~\bibnamefont {Ant{\'o}n}}, \bibinfo {author} {\bibfnamefont
  {J.}~\bibnamefont {Demory}}, \bibinfo {author} {\bibfnamefont
  {C.}~\bibnamefont {G\'{o}mez}}, \bibinfo {author} {\bibfnamefont
  {S.}~\bibnamefont {I.}}, \bibinfo {author} {\bibfnamefont {N.~D.}\
  \bibnamefont {Lanzillotti-Kimura}}, \bibinfo {author} {\bibfnamefont
  {A.}~\bibnamefont {Lema\'{i}tre}}, \bibinfo {author} {\bibfnamefont
  {A.}~\bibnamefont {A.}}, \bibinfo {author} {\bibfnamefont {A.~G.}\
  \bibnamefont {White}}, \bibinfo {author} {\bibfnamefont {L.}~\bibnamefont
  {Lanco}}, \ and\ \bibinfo {author} {\bibfnamefont {P.}~\bibnamefont
  {Senellart}},\ }\href@noop {} {\bibfield  {journal} {\bibinfo  {journal}
  {Nat. Photon.}\ }\textbf {\bibinfo {volume} {10}},\ \bibinfo {pages} {340}
  (\bibinfo {year} {2016})}\BibitemShut {NoStop}%
\bibitem [{\citenamefont {Unsleber}\ \emph {et~al.}(2016)\citenamefont
  {Unsleber}, \citenamefont {He}, \citenamefont {Gerhardt}, \citenamefont
  {Maier}, \citenamefont {Lu}, \citenamefont {Pan}, \citenamefont {Gregersen},
  \citenamefont {Kamp}, \citenamefont {Schneider},\ and\ \citenamefont
  {H{\"o}fling}}]{hofling2016}%
  \BibitemOpen
  \bibfield  {author} {\bibinfo {author} {\bibfnamefont {S.}~\bibnamefont
  {Unsleber}}, \bibinfo {author} {\bibfnamefont {Y.-M.}\ \bibnamefont {He}},
  \bibinfo {author} {\bibfnamefont {S.}~\bibnamefont {Gerhardt}}, \bibinfo
  {author} {\bibfnamefont {S.}~\bibnamefont {Maier}}, \bibinfo {author}
  {\bibfnamefont {C.-Y.}\ \bibnamefont {Lu}}, \bibinfo {author} {\bibfnamefont
  {J.-W.}\ \bibnamefont {Pan}}, \bibinfo {author} {\bibfnamefont
  {N.}~\bibnamefont {Gregersen}}, \bibinfo {author} {\bibfnamefont
  {M.}~\bibnamefont {Kamp}}, \bibinfo {author} {\bibfnamefont {C.}~\bibnamefont
  {Schneider}}, \ and\ \bibinfo {author} {\bibfnamefont {S.}~\bibnamefont
  {H{\"o}fling}},\ }\href@noop {} {\bibfield  {journal} {\bibinfo  {journal}
  {Opt. Express}\ }\textbf {\bibinfo {volume} {24}},\ \bibinfo {pages} {8539}
  (\bibinfo {year} {2016})}\BibitemShut {NoStop}%
\bibitem [{\citenamefont {Wang}\ \emph {et~al.}(2017)\citenamefont {Wang},
  \citenamefont {He}, \citenamefont {Li}, \citenamefont {Su}, \citenamefont
  {Li}, \citenamefont {Huang}, \citenamefont {Ding}, \citenamefont {Chen},
  \citenamefont {Liu}, \citenamefont {Qin}, \citenamefont {Li}, \citenamefont
  {Y.-M.}, \citenamefont {Schneider}, \citenamefont {Kamp}, \citenamefont
  {Peng}, \citenamefont {H{\"o}fling}, \citenamefont {Lu},\ and\ \citenamefont
  {Pan}}]{pan2017}%
  \BibitemOpen
  \bibfield  {author} {\bibinfo {author} {\bibfnamefont {H.}~\bibnamefont
  {Wang}}, \bibinfo {author} {\bibfnamefont {Y.}~\bibnamefont {He}}, \bibinfo
  {author} {\bibfnamefont {Y.-H.}\ \bibnamefont {Li}}, \bibinfo {author}
  {\bibfnamefont {Z.-E.}\ \bibnamefont {Su}}, \bibinfo {author} {\bibfnamefont
  {B.}~\bibnamefont {Li}}, \bibinfo {author} {\bibfnamefont {H.-L.}\
  \bibnamefont {Huang}}, \bibinfo {author} {\bibfnamefont {X.}~\bibnamefont
  {Ding}}, \bibinfo {author} {\bibfnamefont {M.-C.}\ \bibnamefont {Chen}},
  \bibinfo {author} {\bibfnamefont {C.}~\bibnamefont {Liu}}, \bibinfo {author}
  {\bibfnamefont {J.}~\bibnamefont {Qin}}, \bibinfo {author} {\bibfnamefont
  {J.-P.}\ \bibnamefont {Li}}, \bibinfo {author} {\bibfnamefont
  {H.}~\bibnamefont {Y.-M.}}, \bibinfo {author} {\bibfnamefont
  {C.}~\bibnamefont {Schneider}}, \bibinfo {author} {\bibfnamefont
  {M.}~\bibnamefont {Kamp}}, \bibinfo {author} {\bibfnamefont {C.-Z.}\
  \bibnamefont {Peng}}, \bibinfo {author} {\bibfnamefont {S.}~\bibnamefont
  {H{\"o}fling}}, \bibinfo {author} {\bibfnamefont {C.-Y.}\ \bibnamefont {Lu}},
  \ and\ \bibinfo {author} {\bibfnamefont {J.-W.}\ \bibnamefont {Pan}},\
  }\href@noop {} {\bibfield  {journal} {\bibinfo  {journal} {Nat. Photon.}\
  }\textbf {\bibinfo {volume} {11}},\ \bibinfo {pages} {361} (\bibinfo {year}
  {2017})}\BibitemShut {NoStop}%
\bibitem [{\citenamefont {He}\ \emph {et~al.}(2017)\citenamefont {He},
  \citenamefont {Liu}, \citenamefont {Maier}, \citenamefont {Emmerling},
  \citenamefont {Gerhardt}, \citenamefont {Davan\c{c}o}, \citenamefont
  {Srinivasan}, \citenamefont {Schneider},\ and\ \citenamefont
  {H{\"o}fling}}]{hofling2017}%
  \BibitemOpen
  \bibfield  {author} {\bibinfo {author} {\bibfnamefont {Y.-M.}\ \bibnamefont
  {He}}, \bibinfo {author} {\bibfnamefont {J.}~\bibnamefont {Liu}}, \bibinfo
  {author} {\bibfnamefont {S.}~\bibnamefont {Maier}}, \bibinfo {author}
  {\bibfnamefont {M.}~\bibnamefont {Emmerling}}, \bibinfo {author}
  {\bibfnamefont {S.}~\bibnamefont {Gerhardt}}, \bibinfo {author}
  {\bibfnamefont {M.}~\bibnamefont {Davan\c{c}o}}, \bibinfo {author}
  {\bibfnamefont {K.}~\bibnamefont {Srinivasan}}, \bibinfo {author}
  {\bibfnamefont {C.}~\bibnamefont {Schneider}}, \ and\ \bibinfo {author}
  {\bibfnamefont {S.}~\bibnamefont {H{\"o}fling}},\ }\href@noop {} {\bibfield
  {journal} {\bibinfo  {journal} {Optica}\ }\textbf {\bibinfo {volume} {4}},\
  \bibinfo {pages} {802} (\bibinfo {year} {2017})}\BibitemShut {NoStop}%
\bibitem [{\citenamefont {Kir{\v s}ansk{\.e}}\ \emph
  {et~al.}(2017)\citenamefont {Kir{\v s}ansk{\.e}}, \citenamefont
  {Thyrrestrup}, \citenamefont {Daveau}, \citenamefont {Dree\ss{}en},
  \citenamefont {Pregnolato}, \citenamefont {Midolo}, \citenamefont
  {Tighineanu}, \citenamefont {Javadi}, \citenamefont {Stobbe}, \citenamefont
  {Schott}, \citenamefont {Ludwig}, \citenamefont {Wieck}, \citenamefont
  {Park}, \citenamefont {Song}, \citenamefont {Kuhlmann}, \citenamefont
  {S\"ollner}, \citenamefont {L\"obl}, \citenamefont {Warburton},\ and\
  \citenamefont {Lodahl}}]{lodahl2017}%
  \BibitemOpen
  \bibfield  {author} {\bibinfo {author} {\bibfnamefont {G.}~\bibnamefont
  {Kir{\v s}ansk{\.e}}}, \bibinfo {author} {\bibfnamefont {H.}~\bibnamefont
  {Thyrrestrup}}, \bibinfo {author} {\bibfnamefont {R.~S.}\ \bibnamefont
  {Daveau}}, \bibinfo {author} {\bibfnamefont {C.~L.}\ \bibnamefont
  {Dree\ss{}en}}, \bibinfo {author} {\bibfnamefont {T.}~\bibnamefont
  {Pregnolato}}, \bibinfo {author} {\bibfnamefont {L.}~\bibnamefont {Midolo}},
  \bibinfo {author} {\bibfnamefont {P.}~\bibnamefont {Tighineanu}}, \bibinfo
  {author} {\bibfnamefont {A.}~\bibnamefont {Javadi}}, \bibinfo {author}
  {\bibfnamefont {S.}~\bibnamefont {Stobbe}}, \bibinfo {author} {\bibfnamefont
  {R.}~\bibnamefont {Schott}}, \bibinfo {author} {\bibfnamefont
  {A.}~\bibnamefont {Ludwig}}, \bibinfo {author} {\bibfnamefont {A.~D.}\
  \bibnamefont {Wieck}}, \bibinfo {author} {\bibfnamefont {S.~I.}\ \bibnamefont
  {Park}}, \bibinfo {author} {\bibfnamefont {J.~D.}\ \bibnamefont {Song}},
  \bibinfo {author} {\bibfnamefont {A.~V.}\ \bibnamefont {Kuhlmann}}, \bibinfo
  {author} {\bibfnamefont {I.}~\bibnamefont {S\"ollner}}, \bibinfo {author}
  {\bibfnamefont {M.~C.}\ \bibnamefont {L\"obl}}, \bibinfo {author}
  {\bibfnamefont {R.~J.}\ \bibnamefont {Warburton}}, \ and\ \bibinfo {author}
  {\bibfnamefont {P.}~\bibnamefont {Lodahl}},\ }\href@noop {} {\bibfield
  {journal} {\bibinfo  {journal} {Phys. Rev. B}\ }\textbf {\bibinfo {volume}
  {96}},\ \bibinfo {pages} {165306} (\bibinfo {year} {2017})}\BibitemShut
  {NoStop}%
\bibitem [{\citenamefont {Liu}\ \emph {et~al.}(2018{\natexlab{a}})\citenamefont
  {Liu}, \citenamefont {Brash}, \citenamefont {O’Hara}, \citenamefont
  {Martins}, \citenamefont {Phillips}, \citenamefont {Coles}, \citenamefont
  {Royall}, \citenamefont {Clarke}, \citenamefont {Bentham}, \citenamefont
  {Prtljaga}, \citenamefont {Itskevich}, \citenamefont {WIlson},\ and\
  \citenamefont {Fox}}]{fox2018}%
  \BibitemOpen
  \bibfield  {author} {\bibinfo {author} {\bibfnamefont {F.}~\bibnamefont
  {Liu}}, \bibinfo {author} {\bibfnamefont {A.~J.}\ \bibnamefont {Brash}},
  \bibinfo {author} {\bibfnamefont {J.}~\bibnamefont {O’Hara}}, \bibinfo
  {author} {\bibfnamefont {L.~M.}\ \bibnamefont {Martins}}, \bibinfo {author}
  {\bibfnamefont {C.~L.}\ \bibnamefont {Phillips}}, \bibinfo {author}
  {\bibfnamefont {R.~J.}\ \bibnamefont {Coles}}, \bibinfo {author}
  {\bibfnamefont {B.}~\bibnamefont {Royall}}, \bibinfo {author} {\bibfnamefont
  {E.}~\bibnamefont {Clarke}}, \bibinfo {author} {\bibfnamefont
  {C.}~\bibnamefont {Bentham}}, \bibinfo {author} {\bibfnamefont
  {N.}~\bibnamefont {Prtljaga}}, \bibinfo {author} {\bibfnamefont {I.~E.}\
  \bibnamefont {Itskevich}}, \bibinfo {author} {\bibfnamefont {L.~R.}\
  \bibnamefont {WIlson}}, \ and\ \bibinfo {author} {\bibfnamefont {A.~M.}\
  \bibnamefont {Fox}},\ }\href@noop {} {\bibfield  {journal} {\bibinfo
  {journal} {Nat. Nanotechnol.}\ }\textbf {\bibinfo {volume} {13}},\ \bibinfo
  {pages} {835} (\bibinfo {year} {2018}{\natexlab{a}})}\BibitemShut {NoStop}%
\bibitem [{\citenamefont {Wang}\ \emph
  {et~al.}(2019{\natexlab{a}})\citenamefont {Wang}, \citenamefont {He},
  \citenamefont {Chung}, \citenamefont {Hu}, \citenamefont {Yu}, \citenamefont
  {Chen}, \citenamefont {Ding}, \citenamefont {Chen}, \citenamefont {Qin},
  \citenamefont {Yang}, \citenamefont {Liu}, \citenamefont {Duan},
  \citenamefont {Li}, \citenamefont {Gerhardt}, \citenamefont {Winkler},
  \citenamefont {Jurkat}, \citenamefont {Wang}, \citenamefont {Gregersen},
  \citenamefont {Huo}, \citenamefont {Dai}, \citenamefont {Yu}, \citenamefont
  {H{\"o}fling}, \citenamefont {Lu},\ and\ \citenamefont
  {Pan}}]{pan2019elliptical}%
  \BibitemOpen
  \bibfield  {author} {\bibinfo {author} {\bibfnamefont {H.}~\bibnamefont
  {Wang}}, \bibinfo {author} {\bibfnamefont {Y.-M.}\ \bibnamefont {He}},
  \bibinfo {author} {\bibfnamefont {T.-H.}\ \bibnamefont {Chung}}, \bibinfo
  {author} {\bibfnamefont {H.}~\bibnamefont {Hu}}, \bibinfo {author}
  {\bibfnamefont {Y.}~\bibnamefont {Yu}}, \bibinfo {author} {\bibfnamefont
  {S.}~\bibnamefont {Chen}}, \bibinfo {author} {\bibfnamefont {X.}~\bibnamefont
  {Ding}}, \bibinfo {author} {\bibfnamefont {M.-C.}\ \bibnamefont {Chen}},
  \bibinfo {author} {\bibfnamefont {J.}~\bibnamefont {Qin}}, \bibinfo {author}
  {\bibfnamefont {X.}~\bibnamefont {Yang}}, \bibinfo {author} {\bibfnamefont
  {R.-Z.}\ \bibnamefont {Liu}}, \bibinfo {author} {\bibfnamefont {Z.-C.}\
  \bibnamefont {Duan}}, \bibinfo {author} {\bibfnamefont {J.-P.}\ \bibnamefont
  {Li}}, \bibinfo {author} {\bibfnamefont {S.}~\bibnamefont {Gerhardt}},
  \bibinfo {author} {\bibfnamefont {K.}~\bibnamefont {Winkler}}, \bibinfo
  {author} {\bibfnamefont {J.}~\bibnamefont {Jurkat}}, \bibinfo {author}
  {\bibfnamefont {L.-J.}\ \bibnamefont {Wang}}, \bibinfo {author}
  {\bibfnamefont {N.}~\bibnamefont {Gregersen}}, \bibinfo {author}
  {\bibfnamefont {Y.-H.}\ \bibnamefont {Huo}}, \bibinfo {author} {\bibfnamefont
  {Q.}~\bibnamefont {Dai}}, \bibinfo {author} {\bibfnamefont {S.}~\bibnamefont
  {Yu}}, \bibinfo {author} {\bibfnamefont {S.}~\bibnamefont {H{\"o}fling}},
  \bibinfo {author} {\bibfnamefont {C.-Y.}\ \bibnamefont {Lu}}, \ and\ \bibinfo
  {author} {\bibfnamefont {J.-W.}\ \bibnamefont {Pan}},\ }\href@noop {}
  {\bibfield  {journal} {\bibinfo  {journal} {Nat. Photon.}\ }\textbf {\bibinfo
  {volume} {13}},\ \bibinfo {pages} {770} (\bibinfo {year}
  {2019}{\natexlab{a}})}\BibitemShut {NoStop}%
\bibitem [{\citenamefont {Thyrrestrup}\ \emph {et~al.}(2018)\citenamefont
  {Thyrrestrup}, \citenamefont {Kir{\v s}ansk{\.e}}, \citenamefont
  {Le~Jeannic}, \citenamefont {Pregnolato}, \citenamefont {Zhai}, \citenamefont
  {Raahauge}, \citenamefont {Midolo}, \citenamefont {Rotenberg}, \citenamefont
  {Javadi}, \citenamefont {Schott}, \citenamefont {Wieck}, \citenamefont
  {Ludwig}, \citenamefont {L\"{o}bl}, \citenamefont {S\"{o}llner},
  \citenamefont {Warburton},\ and\ \citenamefont {Lodahl}}]{lodahl2018}%
  \BibitemOpen
  \bibfield  {author} {\bibinfo {author} {\bibfnamefont {H.}~\bibnamefont
  {Thyrrestrup}}, \bibinfo {author} {\bibfnamefont {G.}~\bibnamefont {Kir{\v
  s}ansk{\.e}}}, \bibinfo {author} {\bibfnamefont {H.}~\bibnamefont
  {Le~Jeannic}}, \bibinfo {author} {\bibfnamefont {T.}~\bibnamefont
  {Pregnolato}}, \bibinfo {author} {\bibfnamefont {L.}~\bibnamefont {Zhai}},
  \bibinfo {author} {\bibfnamefont {L.}~\bibnamefont {Raahauge}}, \bibinfo
  {author} {\bibfnamefont {L.}~\bibnamefont {Midolo}}, \bibinfo {author}
  {\bibfnamefont {N.}~\bibnamefont {Rotenberg}}, \bibinfo {author}
  {\bibfnamefont {A.}~\bibnamefont {Javadi}}, \bibinfo {author} {\bibfnamefont
  {R.}~\bibnamefont {Schott}}, \bibinfo {author} {\bibfnamefont {A.~D.}\
  \bibnamefont {Wieck}}, \bibinfo {author} {\bibfnamefont {A.}~\bibnamefont
  {Ludwig}}, \bibinfo {author} {\bibfnamefont {M.~C.}\ \bibnamefont
  {L\"{o}bl}}, \bibinfo {author} {\bibfnamefont {I.}~\bibnamefont
  {S\"{o}llner}}, \bibinfo {author} {\bibfnamefont {R.~J.}\ \bibnamefont
  {Warburton}}, \ and\ \bibinfo {author} {\bibfnamefont {P.}~\bibnamefont
  {Lodahl}},\ }\href@noop {} {\bibfield  {journal} {\bibinfo  {journal} {Nano
  Lett.}\ }\textbf {\bibinfo {volume} {18}},\ \bibinfo {pages} {1801} (\bibinfo
  {year} {2018})}\BibitemShut {NoStop}%
\bibitem [{\citenamefont {Uppu}\ \emph {et~al.}(2020)\citenamefont {Uppu},
  \citenamefont {Eriksen}, \citenamefont {Thyrrestrup}, \citenamefont
  {U{\u{g}}urlu}, \citenamefont {Wang}, \citenamefont {Scholz}, \citenamefont
  {Wieck}, \citenamefont {Ludwig}, \citenamefont {L{\"o}bl}, \citenamefont
  {Warburton}, \citenamefont {Lodahl},\ and\ \citenamefont
  {Midolo}}]{Uppu2020}%
  \BibitemOpen
  \bibfield  {author} {\bibinfo {author} {\bibfnamefont {R.}~\bibnamefont
  {Uppu}}, \bibinfo {author} {\bibfnamefont {H.~T.}\ \bibnamefont {Eriksen}},
  \bibinfo {author} {\bibfnamefont {H.}~\bibnamefont {Thyrrestrup}}, \bibinfo
  {author} {\bibfnamefont {A.~D.}\ \bibnamefont {U{\u{g}}urlu}}, \bibinfo
  {author} {\bibfnamefont {Y.}~\bibnamefont {Wang}}, \bibinfo {author}
  {\bibfnamefont {S.}~\bibnamefont {Scholz}}, \bibinfo {author} {\bibfnamefont
  {A.~D.}\ \bibnamefont {Wieck}}, \bibinfo {author} {\bibfnamefont
  {A.}~\bibnamefont {Ludwig}}, \bibinfo {author} {\bibfnamefont {M.~C.}\
  \bibnamefont {L{\"o}bl}}, \bibinfo {author} {\bibfnamefont {R.~J.}\
  \bibnamefont {Warburton}}, \bibinfo {author} {\bibfnamefont {P.}~\bibnamefont
  {Lodahl}}, \ and\ \bibinfo {author} {\bibfnamefont {L.}~\bibnamefont
  {Midolo}},\ }\href@noop {} {\bibfield  {journal} {\bibinfo  {journal}
  {arXiv:2001.10716}\ } (\bibinfo {year} {2020})}\BibitemShut {NoStop}%
\bibitem [{\citenamefont {Lodahl}(2017)}]{lodahl2017quantum}%
  \BibitemOpen
  \bibfield  {author} {\bibinfo {author} {\bibfnamefont {P.}~\bibnamefont
  {Lodahl}},\ }\href@noop {} {\bibfield  {journal} {\bibinfo  {journal}
  {Quantum Sci. Technol.}\ }\textbf {\bibinfo {volume} {3}},\ \bibinfo {pages}
  {013001} (\bibinfo {year} {2017})}\BibitemShut {NoStop}%
\bibitem [{\citenamefont {Preskill}(2018)}]{preskill2018}%
  \BibitemOpen
  \bibfield  {author} {\bibinfo {author} {\bibfnamefont {J.}~\bibnamefont
  {Preskill}},\ }\href@noop {} {\bibfield  {journal} {\bibinfo  {journal}
  {Quantum}\ }\textbf {\bibinfo {volume} {2}},\ \bibinfo {pages} {79} (\bibinfo
  {year} {2018})}\BibitemShut {NoStop}%
\bibitem [{\citenamefont {Aaronson}\ and\ \citenamefont
  {Lijie}(2016)}]{Aaronson2016a}%
  \BibitemOpen
  \bibfield  {author} {\bibinfo {author} {\bibfnamefont {S.}~\bibnamefont
  {Aaronson}}\ and\ \bibinfo {author} {\bibfnamefont {C.}~\bibnamefont
  {Lijie}},\ }\href@noop {} {\bibfield  {journal} {\bibinfo  {journal}
  {arXiv:1612.05903}\ } (\bibinfo {year} {2016})}\BibitemShut {NoStop}%
\bibitem [{\citenamefont {Arute}\ \emph {et~al.}(2019)\citenamefont {Arute},
  \citenamefont {Arya}, \citenamefont {Babbush}, \citenamefont {Bacon},
  \citenamefont {Bardin}, \citenamefont {Barends}, \citenamefont {Biswas},
  \citenamefont {Boixo}, \citenamefont {Brandao}, \citenamefont {Buell},
  \citenamefont {Burkett}, \citenamefont {Chen}, \citenamefont {Chen},
  \citenamefont {Chiaro}, \citenamefont {Collins}, \citenamefont {Courtney},
  \citenamefont {Dunsworth}, \citenamefont {Farhi}, \citenamefont {Foxen},
  \citenamefont {Fowler}, \citenamefont {Gidney}, \citenamefont {Giustina},
  \citenamefont {Graff}, \citenamefont {Guerin}, \citenamefont {Habegger},
  \citenamefont {Harrigan}, \citenamefont {Hartmann}, \citenamefont {Ho},
  \citenamefont {Hoffmann}, \citenamefont {Huang}, \citenamefont {Humble},
  \citenamefont {Isakov}, \citenamefont {Jeffrey}, \citenamefont {Jiang},
  \citenamefont {Kafri}, \citenamefont {Kechedzhi}, \citenamefont {Kelly},
  \citenamefont {Klimov}, \citenamefont {Knysh}, \citenamefont {Korotkov},
  \citenamefont {Kostritsa}, \citenamefont {Landhuis}, \citenamefont
  {Lindmark}, \citenamefont {Lucero}, \citenamefont {Lyakh}, \citenamefont
  {Mandrà}, \citenamefont {McClean}, \citenamefont {McEwen}, \citenamefont
  {Megrant}, \citenamefont {Mi}, \citenamefont {Michielsen}, \citenamefont
  {Mohseni}, \citenamefont {Mutus}, \citenamefont {Naaman}, \citenamefont
  {Neeley}, \citenamefont {Neill}, \citenamefont {Niu}, \citenamefont {Ostby},
  \citenamefont {Petukhov}, \citenamefont {Platt}, \citenamefont {Quintana},
  \citenamefont {Rieffel}, \citenamefont {Roushan}, \citenamefont {Rubin},
  \citenamefont {Sank}, \citenamefont {Satzinger}, \citenamefont {Smelyanskiy},
  \citenamefont {Sung}, \citenamefont {Trevithick}, \citenamefont
  {Vainsencher}, \citenamefont {Villalonga}, \citenamefont {White},
  \citenamefont {Yao}, \citenamefont {Yeh}, \citenamefont {Zalcman},
  \citenamefont {Neven},\ and\ \citenamefont {Martinis}}]{Arute2019}%
  \BibitemOpen
  \bibfield  {author} {\bibinfo {author} {\bibfnamefont {F.}~\bibnamefont
  {Arute}}, \bibinfo {author} {\bibfnamefont {K.}~\bibnamefont {Arya}},
  \bibinfo {author} {\bibfnamefont {R.}~\bibnamefont {Babbush}}, \bibinfo
  {author} {\bibfnamefont {D.}~\bibnamefont {Bacon}}, \bibinfo {author}
  {\bibfnamefont {J.}~\bibnamefont {Bardin}}, \bibinfo {author} {\bibfnamefont
  {R.}~\bibnamefont {Barends}}, \bibinfo {author} {\bibfnamefont
  {R.}~\bibnamefont {Biswas}}, \bibinfo {author} {\bibfnamefont
  {S.}~\bibnamefont {Boixo}}, \bibinfo {author} {\bibfnamefont
  {F.}~\bibnamefont {Brandao}}, \bibinfo {author} {\bibfnamefont
  {D.}~\bibnamefont {Buell}}, \bibinfo {author} {\bibfnamefont
  {B.}~\bibnamefont {Burkett}}, \bibinfo {author} {\bibfnamefont
  {Y.}~\bibnamefont {Chen}}, \bibinfo {author} {\bibfnamefont {J.}~\bibnamefont
  {Chen}}, \bibinfo {author} {\bibfnamefont {B.}~\bibnamefont {Chiaro}},
  \bibinfo {author} {\bibfnamefont {R.}~\bibnamefont {Collins}}, \bibinfo
  {author} {\bibfnamefont {W.}~\bibnamefont {Courtney}}, \bibinfo {author}
  {\bibfnamefont {A.}~\bibnamefont {Dunsworth}}, \bibinfo {author}
  {\bibfnamefont {E.}~\bibnamefont {Farhi}}, \bibinfo {author} {\bibfnamefont
  {B.}~\bibnamefont {Foxen}}, \bibinfo {author} {\bibfnamefont
  {A.}~\bibnamefont {Fowler}}, \bibinfo {author} {\bibfnamefont {C.~M.}\
  \bibnamefont {Gidney}}, \bibinfo {author} {\bibfnamefont {M.}~\bibnamefont
  {Giustina}}, \bibinfo {author} {\bibfnamefont {R.}~\bibnamefont {Graff}},
  \bibinfo {author} {\bibfnamefont {K.}~\bibnamefont {Guerin}}, \bibinfo
  {author} {\bibfnamefont {S.}~\bibnamefont {Habegger}}, \bibinfo {author}
  {\bibfnamefont {M.}~\bibnamefont {Harrigan}}, \bibinfo {author}
  {\bibfnamefont {M.}~\bibnamefont {Hartmann}}, \bibinfo {author}
  {\bibfnamefont {A.}~\bibnamefont {Ho}}, \bibinfo {author} {\bibfnamefont
  {M.~R.}\ \bibnamefont {Hoffmann}}, \bibinfo {author} {\bibfnamefont
  {T.}~\bibnamefont {Huang}}, \bibinfo {author} {\bibfnamefont
  {T.}~\bibnamefont {Humble}}, \bibinfo {author} {\bibfnamefont
  {S.}~\bibnamefont {Isakov}}, \bibinfo {author} {\bibfnamefont
  {E.}~\bibnamefont {Jeffrey}}, \bibinfo {author} {\bibfnamefont
  {Z.}~\bibnamefont {Jiang}}, \bibinfo {author} {\bibfnamefont
  {D.}~\bibnamefont {Kafri}}, \bibinfo {author} {\bibfnamefont
  {K.}~\bibnamefont {Kechedzhi}}, \bibinfo {author} {\bibfnamefont
  {J.}~\bibnamefont {Kelly}}, \bibinfo {author} {\bibfnamefont
  {P.}~\bibnamefont {Klimov}}, \bibinfo {author} {\bibfnamefont
  {S.}~\bibnamefont {Knysh}}, \bibinfo {author} {\bibfnamefont
  {A.}~\bibnamefont {Korotkov}}, \bibinfo {author} {\bibfnamefont
  {F.}~\bibnamefont {Kostritsa}}, \bibinfo {author} {\bibfnamefont
  {D.}~\bibnamefont {Landhuis}}, \bibinfo {author} {\bibfnamefont
  {M.}~\bibnamefont {Lindmark}}, \bibinfo {author} {\bibfnamefont
  {E.}~\bibnamefont {Lucero}}, \bibinfo {author} {\bibfnamefont
  {D.}~\bibnamefont {Lyakh}}, \bibinfo {author} {\bibfnamefont
  {S.}~\bibnamefont {Mandrà}}, \bibinfo {author} {\bibfnamefont {J.~R.}\
  \bibnamefont {McClean}}, \bibinfo {author} {\bibfnamefont {M.}~\bibnamefont
  {McEwen}}, \bibinfo {author} {\bibfnamefont {A.}~\bibnamefont {Megrant}},
  \bibinfo {author} {\bibfnamefont {X.}~\bibnamefont {Mi}}, \bibinfo {author}
  {\bibfnamefont {K.}~\bibnamefont {Michielsen}}, \bibinfo {author}
  {\bibfnamefont {M.}~\bibnamefont {Mohseni}}, \bibinfo {author} {\bibfnamefont
  {J.}~\bibnamefont {Mutus}}, \bibinfo {author} {\bibfnamefont
  {O.}~\bibnamefont {Naaman}}, \bibinfo {author} {\bibfnamefont
  {M.}~\bibnamefont {Neeley}}, \bibinfo {author} {\bibfnamefont
  {C.}~\bibnamefont {Neill}}, \bibinfo {author} {\bibfnamefont {M.~Y.}\
  \bibnamefont {Niu}}, \bibinfo {author} {\bibfnamefont {E.}~\bibnamefont
  {Ostby}}, \bibinfo {author} {\bibfnamefont {A.}~\bibnamefont {Petukhov}},
  \bibinfo {author} {\bibfnamefont {J.}~\bibnamefont {Platt}}, \bibinfo
  {author} {\bibfnamefont {C.}~\bibnamefont {Quintana}}, \bibinfo {author}
  {\bibfnamefont {E.~G.}\ \bibnamefont {Rieffel}}, \bibinfo {author}
  {\bibfnamefont {P.}~\bibnamefont {Roushan}}, \bibinfo {author} {\bibfnamefont
  {N.}~\bibnamefont {Rubin}}, \bibinfo {author} {\bibfnamefont
  {D.}~\bibnamefont {Sank}}, \bibinfo {author} {\bibfnamefont {K.~J.}\
  \bibnamefont {Satzinger}}, \bibinfo {author} {\bibfnamefont {V.}~\bibnamefont
  {Smelyanskiy}}, \bibinfo {author} {\bibfnamefont {K.~J.}\ \bibnamefont
  {Sung}}, \bibinfo {author} {\bibfnamefont {M.}~\bibnamefont {Trevithick}},
  \bibinfo {author} {\bibfnamefont {A.}~\bibnamefont {Vainsencher}}, \bibinfo
  {author} {\bibfnamefont {B.}~\bibnamefont {Villalonga}}, \bibinfo {author}
  {\bibfnamefont {T.}~\bibnamefont {White}}, \bibinfo {author} {\bibfnamefont
  {Z.~J.}\ \bibnamefont {Yao}}, \bibinfo {author} {\bibfnamefont
  {P.}~\bibnamefont {Yeh}}, \bibinfo {author} {\bibfnamefont {A.}~\bibnamefont
  {Zalcman}}, \bibinfo {author} {\bibfnamefont {H.}~\bibnamefont {Neven}}, \
  and\ \bibinfo {author} {\bibfnamefont {J.}~\bibnamefont {Martinis}},\
  }\href@noop {} {\bibfield  {journal} {\bibinfo  {journal} {Nature}\ }\textbf
  {\bibinfo {volume} {574}},\ \bibinfo {pages} {505–510} (\bibinfo {year}
  {2019})}\BibitemShut {NoStop}%
\bibitem [{\citenamefont {Wang}\ \emph
  {et~al.}(2019{\natexlab{b}})\citenamefont {Wang}, \citenamefont {Qin},
  \citenamefont {Ding}, \citenamefont {Chen}, \citenamefont {Chen},
  \citenamefont {You}, \citenamefont {He}, \citenamefont {Jiang}, \citenamefont
  {Wang}, \citenamefont {You}, \citenamefont {Renema}, \citenamefont
  {H{\"o}fling}, \citenamefont {Lu},\ and\ \citenamefont {Pan}}]{pan2019boson}%
  \BibitemOpen
  \bibfield  {author} {\bibinfo {author} {\bibfnamefont {H.}~\bibnamefont
  {Wang}}, \bibinfo {author} {\bibfnamefont {J.}~\bibnamefont {Qin}}, \bibinfo
  {author} {\bibfnamefont {X.}~\bibnamefont {Ding}}, \bibinfo {author}
  {\bibfnamefont {M.-C.}\ \bibnamefont {Chen}}, \bibinfo {author}
  {\bibfnamefont {S.}~\bibnamefont {Chen}}, \bibinfo {author} {\bibfnamefont
  {X.}~\bibnamefont {You}}, \bibinfo {author} {\bibfnamefont {Y.-M.}\
  \bibnamefont {He}}, \bibinfo {author} {\bibfnamefont {X.}~\bibnamefont
  {Jiang}}, \bibinfo {author} {\bibfnamefont {Z.}~\bibnamefont {Wang}},
  \bibinfo {author} {\bibfnamefont {L.}~\bibnamefont {You}}, \bibinfo {author}
  {\bibfnamefont {J.~J.}\ \bibnamefont {Renema}}, \bibinfo {author}
  {\bibfnamefont {S.}~\bibnamefont {H{\"o}fling}}, \bibinfo {author}
  {\bibfnamefont {C.-Y.}\ \bibnamefont {Lu}}, \ and\ \bibinfo {author}
  {\bibfnamefont {J.-W.}\ \bibnamefont {Pan}},\ }\href@noop {} {\bibfield
  {journal} {\bibinfo  {journal} {Phys. Rev. Lett.}\ }\textbf {\bibinfo
  {volume} {123}},\ \bibinfo {pages} {250503} (\bibinfo {year}
  {2019}{\natexlab{b}})}\BibitemShut {NoStop}%
\bibitem [{\citenamefont {Kuhlmann}\ \emph {et~al.}(2013)\citenamefont
  {Kuhlmann}, \citenamefont {Houel}, \citenamefont {Ludwig}, \citenamefont
  {Greuter}, \citenamefont {Reuter}, \citenamefont {Wieck}, \citenamefont
  {Poggio},\ and\ \citenamefont {Warburton}}]{kuhlmann2013}%
  \BibitemOpen
  \bibfield  {author} {\bibinfo {author} {\bibfnamefont {A.~V.}\ \bibnamefont
  {Kuhlmann}}, \bibinfo {author} {\bibfnamefont {J.}~\bibnamefont {Houel}},
  \bibinfo {author} {\bibfnamefont {A.}~\bibnamefont {Ludwig}}, \bibinfo
  {author} {\bibfnamefont {L.}~\bibnamefont {Greuter}}, \bibinfo {author}
  {\bibfnamefont {D.}~\bibnamefont {Reuter}}, \bibinfo {author} {\bibfnamefont
  {A.~D.}\ \bibnamefont {Wieck}}, \bibinfo {author} {\bibfnamefont
  {M.}~\bibnamefont {Poggio}}, \ and\ \bibinfo {author} {\bibfnamefont {R.~J.}\
  \bibnamefont {Warburton}},\ }\href@noop {} {\bibfield  {journal} {\bibinfo
  {journal} {Nat. Phys.}\ }\textbf {\bibinfo {volume} {9}},\ \bibinfo {pages}
  {570} (\bibinfo {year} {2013})}\BibitemShut {NoStop}%
\bibitem [{\citenamefont {Hummel}\ \emph {et~al.}(2019)\citenamefont {Hummel},
  \citenamefont {Ouellet-Plamondon}, \citenamefont {Ugur}, \citenamefont
  {Kulkova}, \citenamefont {Lund-Hansen}, \citenamefont {Broome}, \citenamefont
  {Uppu},\ and\ \citenamefont {Lodahl}}]{lodahl2019demux}%
  \BibitemOpen
  \bibfield  {author} {\bibinfo {author} {\bibfnamefont {T.}~\bibnamefont
  {Hummel}}, \bibinfo {author} {\bibfnamefont {C.}~\bibnamefont
  {Ouellet-Plamondon}}, \bibinfo {author} {\bibfnamefont {E.}~\bibnamefont
  {Ugur}}, \bibinfo {author} {\bibfnamefont {I.}~\bibnamefont {Kulkova}},
  \bibinfo {author} {\bibfnamefont {T.}~\bibnamefont {Lund-Hansen}}, \bibinfo
  {author} {\bibfnamefont {M.~A.}\ \bibnamefont {Broome}}, \bibinfo {author}
  {\bibfnamefont {R.}~\bibnamefont {Uppu}}, \ and\ \bibinfo {author}
  {\bibfnamefont {P.}~\bibnamefont {Lodahl}},\ }\href@noop {} {\bibfield
  {journal} {\bibinfo  {journal} {Appl. Phys. Lett.}\ }\textbf {\bibinfo
  {volume} {115}},\ \bibinfo {pages} {021102} (\bibinfo {year}
  {2019})}\BibitemShut {NoStop}%
\bibitem [{\citenamefont {Lund}\ \emph {et~al.}(2017)\citenamefont {Lund},
  \citenamefont {Bremner},\ and\ \citenamefont {Ralph}}]{ralph2017}%
  \BibitemOpen
  \bibfield  {author} {\bibinfo {author} {\bibfnamefont {A.}~\bibnamefont
  {Lund}}, \bibinfo {author} {\bibfnamefont {M.~J.}\ \bibnamefont {Bremner}}, \
  and\ \bibinfo {author} {\bibfnamefont {T.}~\bibnamefont {Ralph}},\
  }\href@noop {} {\bibfield  {journal} {\bibinfo  {journal} {npj Quantum Inf.}\
  }\textbf {\bibinfo {volume} {3}},\ \bibinfo {pages} {15} (\bibinfo {year}
  {2017})}\BibitemShut {NoStop}%
\bibitem [{\citenamefont {Wang}\ \emph {et~al.}(2016)\citenamefont {Wang},
  \citenamefont {Duan}, \citenamefont {Li}, \citenamefont {Chen}, \citenamefont
  {Li}, \citenamefont {He}, \citenamefont {Chen}, \citenamefont {He},
  \citenamefont {Ding}, \citenamefont {Peng}, \citenamefont {Schneider},
  \citenamefont {Kamp}, \citenamefont {H{\"o}fling}, \citenamefont {Lu},\ and\
  \citenamefont {Pan}}]{pan2016}%
  \BibitemOpen
  \bibfield  {author} {\bibinfo {author} {\bibfnamefont {H.}~\bibnamefont
  {Wang}}, \bibinfo {author} {\bibfnamefont {Z.-C.}\ \bibnamefont {Duan}},
  \bibinfo {author} {\bibfnamefont {Y.-H.}\ \bibnamefont {Li}}, \bibinfo
  {author} {\bibfnamefont {S.}~\bibnamefont {Chen}}, \bibinfo {author}
  {\bibfnamefont {J.-P.}\ \bibnamefont {Li}}, \bibinfo {author} {\bibfnamefont
  {Y.-M.}\ \bibnamefont {He}}, \bibinfo {author} {\bibfnamefont {M.-C.}\
  \bibnamefont {Chen}}, \bibinfo {author} {\bibfnamefont {Y.}~\bibnamefont
  {He}}, \bibinfo {author} {\bibfnamefont {X.}~\bibnamefont {Ding}}, \bibinfo
  {author} {\bibfnamefont {C.-Z.}\ \bibnamefont {Peng}}, \bibinfo {author}
  {\bibfnamefont {C.}~\bibnamefont {Schneider}}, \bibinfo {author}
  {\bibfnamefont {M.}~\bibnamefont {Kamp}}, \bibinfo {author} {\bibfnamefont
  {S.}~\bibnamefont {H{\"o}fling}}, \bibinfo {author} {\bibfnamefont {C.-Y.}\
  \bibnamefont {Lu}}, \ and\ \bibinfo {author} {\bibfnamefont {J.-W.}\
  \bibnamefont {Pan}},\ }\href@noop {} {\bibfield  {journal} {\bibinfo
  {journal} {Phys. Rev. Lett.}\ }\textbf {\bibinfo {volume} {116}},\ \bibinfo
  {pages} {213601} (\bibinfo {year} {2016})}\BibitemShut {NoStop}%
\bibitem [{\citenamefont {Kuhlmann}\ \emph {et~al.}(2015)\citenamefont
  {Kuhlmann}, \citenamefont {Prechtel}, \citenamefont {Houel}, \citenamefont
  {Ludwig}, \citenamefont {Reuter}, \citenamefont {Wieck},\ and\ \citenamefont
  {Warburton}}]{kuhlmann2015}%
  \BibitemOpen
  \bibfield  {author} {\bibinfo {author} {\bibfnamefont {A.~V.}\ \bibnamefont
  {Kuhlmann}}, \bibinfo {author} {\bibfnamefont {J.~H.}\ \bibnamefont
  {Prechtel}}, \bibinfo {author} {\bibfnamefont {J.}~\bibnamefont {Houel}},
  \bibinfo {author} {\bibfnamefont {A.}~\bibnamefont {Ludwig}}, \bibinfo
  {author} {\bibfnamefont {D.}~\bibnamefont {Reuter}}, \bibinfo {author}
  {\bibfnamefont {A.~D.}\ \bibnamefont {Wieck}}, \ and\ \bibinfo {author}
  {\bibfnamefont {R.~J.}\ \bibnamefont {Warburton}},\ }\href@noop {} {\bibfield
   {journal} {\bibinfo  {journal} {Nat. Commun.}\ }\textbf {\bibinfo {volume}
  {6}},\ \bibinfo {pages} {8204} (\bibinfo {year} {2015})}\BibitemShut
  {NoStop}%
\bibitem [{\citenamefont {Warburton}(2013)}]{warburton2013}%
  \BibitemOpen
  \bibfield  {author} {\bibinfo {author} {\bibfnamefont {R.~J.}\ \bibnamefont
  {Warburton}},\ }\href@noop {} {\bibfield  {journal} {\bibinfo  {journal}
  {Nat. Mater.}\ }\textbf {\bibinfo {volume} {12}},\ \bibinfo {pages} {483}
  (\bibinfo {year} {2013})}\BibitemShut {NoStop}%
\bibitem [{\citenamefont {Liu}\ \emph {et~al.}(2018{\natexlab{b}})\citenamefont
  {Liu}, \citenamefont {Konthasinghe}, \citenamefont {Davan\c{c}o},
  \citenamefont {Lawall}, \citenamefont {Anant}, \citenamefont {Verma},
  \citenamefont {Mirin}, \citenamefont {Nam}, \citenamefont {Song},
  \citenamefont {Ma}, \citenamefont {Chen}, \citenamefont {Ni}, \citenamefont
  {Niu},\ and\ \citenamefont {Srinivasan}}]{srinivasan2018}%
  \BibitemOpen
  \bibfield  {author} {\bibinfo {author} {\bibfnamefont {J.}~\bibnamefont
  {Liu}}, \bibinfo {author} {\bibfnamefont {K.}~\bibnamefont {Konthasinghe}},
  \bibinfo {author} {\bibfnamefont {M.}~\bibnamefont {Davan\c{c}o}}, \bibinfo
  {author} {\bibfnamefont {J.}~\bibnamefont {Lawall}}, \bibinfo {author}
  {\bibfnamefont {V.}~\bibnamefont {Anant}}, \bibinfo {author} {\bibfnamefont
  {V.}~\bibnamefont {Verma}}, \bibinfo {author} {\bibfnamefont
  {R.}~\bibnamefont {Mirin}}, \bibinfo {author} {\bibfnamefont {S.~W.}\
  \bibnamefont {Nam}}, \bibinfo {author} {\bibfnamefont {J.~D.}\ \bibnamefont
  {Song}}, \bibinfo {author} {\bibfnamefont {B.}~\bibnamefont {Ma}}, \bibinfo
  {author} {\bibfnamefont {Z.~S.}\ \bibnamefont {Chen}}, \bibinfo {author}
  {\bibfnamefont {H.~Q.}\ \bibnamefont {Ni}}, \bibinfo {author} {\bibfnamefont
  {Z.~C.}\ \bibnamefont {Niu}}, \ and\ \bibinfo {author} {\bibfnamefont
  {K.}~\bibnamefont {Srinivasan}},\ }\href@noop {} {\bibfield  {journal}
  {\bibinfo  {journal} {Phys. Rev. Applied}\ }\textbf {\bibinfo {volume} {9}},\
  \bibinfo {pages} {064019} (\bibinfo {year} {2018}{\natexlab{b}})}\BibitemShut
  {NoStop}%
\bibitem [{\citenamefont {Das}\ \emph {et~al.}(2019)\citenamefont {Das},
  \citenamefont {Zhai}, \citenamefont {{\v{C}}epulskovskis}, \citenamefont
  {Javadi}, \citenamefont {Mahmoodian}, \citenamefont {Lodahl},\ and\
  \citenamefont {S{\o}rensen}}]{Sumanta2019}%
  \BibitemOpen
  \bibfield  {author} {\bibinfo {author} {\bibfnamefont {S.}~\bibnamefont
  {Das}}, \bibinfo {author} {\bibfnamefont {L.}~\bibnamefont {Zhai}}, \bibinfo
  {author} {\bibfnamefont {M.}~\bibnamefont {{\v{C}}epulskovskis}}, \bibinfo
  {author} {\bibfnamefont {A.}~\bibnamefont {Javadi}}, \bibinfo {author}
  {\bibfnamefont {S.}~\bibnamefont {Mahmoodian}}, \bibinfo {author}
  {\bibfnamefont {P.}~\bibnamefont {Lodahl}}, \ and\ \bibinfo {author}
  {\bibfnamefont {A.~S.}\ \bibnamefont {S{\o}rensen}},\ }\href@noop {}
  {\bibfield  {journal} {\bibinfo  {journal} {arXiv:1912.08303}\ } (\bibinfo
  {year} {2019})}\BibitemShut {NoStop}%
\bibitem [{\citenamefont {Schweickert}\ \emph {et~al.}(2018)\citenamefont
  {Schweickert}, \citenamefont {Jöns}, \citenamefont {Zeuner}, \citenamefont
  {Covre~da Silva}, \citenamefont {Huang}, \citenamefont {Lettner},
  \citenamefont {Reindl}, \citenamefont {Zichi}, \citenamefont {Trotta},
  \citenamefont {Rastelli},\ and\ \citenamefont {Zwiller}}]{zwiller2018}%
  \BibitemOpen
  \bibfield  {author} {\bibinfo {author} {\bibfnamefont {L.}~\bibnamefont
  {Schweickert}}, \bibinfo {author} {\bibfnamefont {K.~D.}\ \bibnamefont
  {Jöns}}, \bibinfo {author} {\bibfnamefont {K.~D.}\ \bibnamefont {Zeuner}},
  \bibinfo {author} {\bibfnamefont {S.~F.}\ \bibnamefont {Covre~da Silva}},
  \bibinfo {author} {\bibfnamefont {H.}~\bibnamefont {Huang}}, \bibinfo
  {author} {\bibfnamefont {T.}~\bibnamefont {Lettner}}, \bibinfo {author}
  {\bibfnamefont {M.}~\bibnamefont {Reindl}}, \bibinfo {author} {\bibfnamefont
  {J.}~\bibnamefont {Zichi}}, \bibinfo {author} {\bibfnamefont
  {R.}~\bibnamefont {Trotta}}, \bibinfo {author} {\bibfnamefont
  {A.}~\bibnamefont {Rastelli}}, \ and\ \bibinfo {author} {\bibfnamefont
  {V.}~\bibnamefont {Zwiller}},\ }\href@noop {} {\bibfield  {journal} {\bibinfo
   {journal} {Appl. Phys. Lett.}\ }\textbf {\bibinfo {volume} {112}},\ \bibinfo
  {pages} {093106} (\bibinfo {year} {2018})}\BibitemShut {NoStop}%
\bibitem [{\citenamefont {Tighineanu}\ \emph {et~al.}(2018)\citenamefont
  {Tighineanu}, \citenamefont {Dreessen}, \citenamefont {Flindt}, \citenamefont
  {Lodahl},\ and\ \citenamefont {S{\o}rensen}}]{sorensen2018}%
  \BibitemOpen
  \bibfield  {author} {\bibinfo {author} {\bibfnamefont {P.}~\bibnamefont
  {Tighineanu}}, \bibinfo {author} {\bibfnamefont {C.~L.}\ \bibnamefont
  {Dreessen}}, \bibinfo {author} {\bibfnamefont {C.}~\bibnamefont {Flindt}},
  \bibinfo {author} {\bibfnamefont {P.}~\bibnamefont {Lodahl}}, \ and\ \bibinfo
  {author} {\bibfnamefont {A.~S.}\ \bibnamefont {S{\o}rensen}},\ }\href@noop {}
  {\bibfield  {journal} {\bibinfo  {journal} {Phys. Rev. Lett.}\ }\textbf
  {\bibinfo {volume} {120}},\ \bibinfo {pages} {257401} (\bibinfo {year}
  {2018})}\BibitemShut {NoStop}%
\bibitem [{\citenamefont {Renema}\ \emph {et~al.}(2018)\citenamefont {Renema},
  \citenamefont {Menssen}, \citenamefont {Clements}, \citenamefont {Triginer},
  \citenamefont {Kolthammer},\ and\ \citenamefont {Walmsley}}]{Renema2018}%
  \BibitemOpen
  \bibfield  {author} {\bibinfo {author} {\bibfnamefont {J.~J.}\ \bibnamefont
  {Renema}}, \bibinfo {author} {\bibfnamefont {A.}~\bibnamefont {Menssen}},
  \bibinfo {author} {\bibfnamefont {W.~R.}\ \bibnamefont {Clements}}, \bibinfo
  {author} {\bibfnamefont {G.}~\bibnamefont {Triginer}}, \bibinfo {author}
  {\bibfnamefont {W.~S.}\ \bibnamefont {Kolthammer}}, \ and\ \bibinfo {author}
  {\bibfnamefont {I.~A.}\ \bibnamefont {Walmsley}},\ }\href@noop {} {\bibfield
  {journal} {\bibinfo  {journal} {Phys. Rev. Lett.}\ }\textbf {\bibinfo
  {volume} {120}},\ \bibinfo {pages} {220502} (\bibinfo {year}
  {2018})}\BibitemShut {NoStop}%
\bibitem [{\citenamefont {Tichy}(2015)}]{Tichy2015}%
  \BibitemOpen
  \bibfield  {author} {\bibinfo {author} {\bibfnamefont {M.~C.}\ \bibnamefont
  {Tichy}},\ }\href@noop {} {\bibfield  {journal} {\bibinfo  {journal} {Phys.
  Rev. A}\ }\textbf {\bibinfo {volume} {91}},\ \bibinfo {pages} {022316}
  (\bibinfo {year} {2015})}\BibitemShut {NoStop}%
\bibitem [{\citenamefont {Shchesnovich}(2015)}]{Shchesnovich2015}%
  \BibitemOpen
  \bibfield  {author} {\bibinfo {author} {\bibfnamefont {V.~S.}\ \bibnamefont
  {Shchesnovich}},\ }\href@noop {} {\bibfield  {journal} {\bibinfo  {journal}
  {Phys. Rev. A}\ }\textbf {\bibinfo {volume} {91}},\ \bibinfo {pages} {063842}
  (\bibinfo {year} {2015})}\BibitemShut {NoStop}%
\bibitem [{\citenamefont {Aaronson}\ and\ \citenamefont
  {Brod}(2016)}]{Aaronson2016}%
  \BibitemOpen
  \bibfield  {author} {\bibinfo {author} {\bibfnamefont {S.}~\bibnamefont
  {Aaronson}}\ and\ \bibinfo {author} {\bibfnamefont {D.~J.}\ \bibnamefont
  {Brod}},\ }\href@noop {} {\bibfield  {journal} {\bibinfo  {journal} {Phys.
  Rev. A}\ }\textbf {\bibinfo {volume} {93}},\ \bibinfo {pages} {012335}
  (\bibinfo {year} {2016})}\BibitemShut {NoStop}%
\bibitem [{\citenamefont {L{\"o}bl}\ \emph {et~al.}(2019)\citenamefont
  {L{\"o}bl}, \citenamefont {Scholz}, \citenamefont {S{\"o}llner},
  \citenamefont {Ritzmann}, \citenamefont {Denneulin}, \citenamefont
  {Kov{\'a}cs}, \citenamefont {Kardyna{\l}}, \citenamefont {Wieck},
  \citenamefont {Ludwig},\ and\ \citenamefont {Warburton}}]{Lobl2019}%
  \BibitemOpen
  \bibfield  {author} {\bibinfo {author} {\bibfnamefont {M.~C.}\ \bibnamefont
  {L{\"o}bl}}, \bibinfo {author} {\bibfnamefont {S.}~\bibnamefont {Scholz}},
  \bibinfo {author} {\bibfnamefont {I.}~\bibnamefont {S{\"o}llner}}, \bibinfo
  {author} {\bibfnamefont {J.}~\bibnamefont {Ritzmann}}, \bibinfo {author}
  {\bibfnamefont {T.}~\bibnamefont {Denneulin}}, \bibinfo {author}
  {\bibfnamefont {A.}~\bibnamefont {Kov{\'a}cs}}, \bibinfo {author}
  {\bibfnamefont {B.~E.}\ \bibnamefont {Kardyna{\l}}}, \bibinfo {author}
  {\bibfnamefont {A.~D.}\ \bibnamefont {Wieck}}, \bibinfo {author}
  {\bibfnamefont {A.}~\bibnamefont {Ludwig}}, \ and\ \bibinfo {author}
  {\bibfnamefont {R.~J.}\ \bibnamefont {Warburton}},\ }\href@noop {} {\bibfield
   {journal} {\bibinfo  {journal} {Commun. Phys.}\ }\textbf {\bibinfo {volume}
  {2}},\ \bibinfo {pages} {93} (\bibinfo {year} {2019})}\BibitemShut {NoStop}%
\bibitem [{\citenamefont {Midolo}\ \emph {et~al.}(2015)\citenamefont {Midolo},
  \citenamefont {Pregnolato}, \citenamefont {Kir{\v{s}}ansk{\.e}},\ and\
  \citenamefont {Stobbe}}]{midolo_soft_2015}%
  \BibitemOpen
  \bibfield  {author} {\bibinfo {author} {\bibfnamefont {L.}~\bibnamefont
  {Midolo}}, \bibinfo {author} {\bibfnamefont {T.}~\bibnamefont {Pregnolato}},
  \bibinfo {author} {\bibfnamefont {G.}~\bibnamefont {Kir{\v{s}}ansk{\.e}}}, \
  and\ \bibinfo {author} {\bibfnamefont {S.}~\bibnamefont {Stobbe}},\
  }\href@noop {} {\bibfield  {journal} {\bibinfo  {journal} {Nanotechnology}\
  }\textbf {\bibinfo {volume} {26}},\ \bibinfo {pages} {484002} (\bibinfo
  {year} {2015})}\BibitemShut {NoStop}%
\bibitem [{\citenamefont {Santori}\ \emph {et~al.}(2002)\citenamefont
  {Santori}, \citenamefont {Fattal}, \citenamefont {Vu{\v{c}}kovi{\'c}},
  \citenamefont {Solomon},\ and\ \citenamefont {Yamamoto}}]{santori2002}%
  \BibitemOpen
  \bibfield  {author} {\bibinfo {author} {\bibfnamefont {C.}~\bibnamefont
  {Santori}}, \bibinfo {author} {\bibfnamefont {D.}~\bibnamefont {Fattal}},
  \bibinfo {author} {\bibfnamefont {J.}~\bibnamefont {Vu{\v{c}}kovi{\'c}}},
  \bibinfo {author} {\bibfnamefont {G.~S.}\ \bibnamefont {Solomon}}, \ and\
  \bibinfo {author} {\bibfnamefont {Y.}~\bibnamefont {Yamamoto}},\ }\href@noop
  {} {\bibfield  {journal} {\bibinfo  {journal} {nature}\ }\textbf {\bibinfo
  {volume} {419}},\ \bibinfo {pages} {594} (\bibinfo {year}
  {2002})}\BibitemShut {NoStop}%
\bibitem [{\citenamefont {Ruiz}\ \emph {et~al.}(2020)\citenamefont {Ruiz},
  \citenamefont {S{\o}rensen} \emph {et~al.}}]{Eva2020}%
  \BibitemOpen
  \bibfield  {author} {\bibinfo {author} {\bibfnamefont {E.~M.~G.}\
  \bibnamefont {Ruiz}}, \bibinfo {author} {\bibfnamefont {A.~S.}\ \bibnamefont
  {S{\o}rensen}},  \emph {et~al.},\ }\href@noop {} {\bibfield  {journal}
  {\bibinfo  {journal} {In preparation}\ } (\bibinfo {year}
  {2020})}\BibitemShut {NoStop}%
\bibitem [{\citenamefont {Javadi}\ \emph {et~al.}(2015)\citenamefont {Javadi},
  \citenamefont {S{\"o}llner}, \citenamefont {Arcari}, \citenamefont {Hansen},
  \citenamefont {Midolo}, \citenamefont {Mahmoodian}, \citenamefont
  {Kir{\v{s}}ansk{\.e}}, \citenamefont {Pregnolato}, \citenamefont {Lee},
  \citenamefont {Song}, \citenamefont {Stobbe},\ and\ \citenamefont
  {Lodahl}}]{javadi2015}%
  \BibitemOpen
  \bibfield  {author} {\bibinfo {author} {\bibfnamefont {A.}~\bibnamefont
  {Javadi}}, \bibinfo {author} {\bibfnamefont {I.}~\bibnamefont {S{\"o}llner}},
  \bibinfo {author} {\bibfnamefont {M.}~\bibnamefont {Arcari}}, \bibinfo
  {author} {\bibfnamefont {S.~L.}\ \bibnamefont {Hansen}}, \bibinfo {author}
  {\bibfnamefont {L.}~\bibnamefont {Midolo}}, \bibinfo {author} {\bibfnamefont
  {S.}~\bibnamefont {Mahmoodian}}, \bibinfo {author} {\bibfnamefont
  {G.}~\bibnamefont {Kir{\v{s}}ansk{\.e}}}, \bibinfo {author} {\bibfnamefont
  {T.}~\bibnamefont {Pregnolato}}, \bibinfo {author} {\bibfnamefont
  {E.}~\bibnamefont {Lee}}, \bibinfo {author} {\bibfnamefont {J.}~\bibnamefont
  {Song}}, \bibinfo {author} {\bibfnamefont {S.}~\bibnamefont {Stobbe}}, \ and\
  \bibinfo {author} {\bibfnamefont {P.}~\bibnamefont {Lodahl}},\ }\href@noop {}
  {\bibfield  {journal} {\bibinfo  {journal} {Nat. Commun.}\ }\textbf {\bibinfo
  {volume} {6}},\ \bibinfo {pages} {8655} (\bibinfo {year} {2015})}\BibitemShut
  {NoStop}%
\bibitem [{\citenamefont {Kako}\ \emph {et~al.}(2006)\citenamefont {Kako},
  \citenamefont {Santori}, \citenamefont {Hoshino}, \citenamefont
  {G{\"o}tzinger}, \citenamefont {Yamamoto},\ and\ \citenamefont
  {Arakawa}}]{kako2006}%
  \BibitemOpen
  \bibfield  {author} {\bibinfo {author} {\bibfnamefont {S.}~\bibnamefont
  {Kako}}, \bibinfo {author} {\bibfnamefont {C.}~\bibnamefont {Santori}},
  \bibinfo {author} {\bibfnamefont {K.}~\bibnamefont {Hoshino}}, \bibinfo
  {author} {\bibfnamefont {S.}~\bibnamefont {G{\"o}tzinger}}, \bibinfo {author}
  {\bibfnamefont {Y.}~\bibnamefont {Yamamoto}}, \ and\ \bibinfo {author}
  {\bibfnamefont {Y.}~\bibnamefont {Arakawa}},\ }\href@noop {} {\bibfield
  {journal} {\bibinfo  {journal} {Nat. Mater.}\ }\textbf {\bibinfo {volume}
  {5}},\ \bibinfo {pages} {887} (\bibinfo {year} {2006})}\BibitemShut {NoStop}%
\bibitem [{\citenamefont {Johansen}\ \emph {et~al.}(2010)\citenamefont
  {Johansen}, \citenamefont {Julsgaard}, \citenamefont {Stobbe}, \citenamefont
  {Hvam},\ and\ \citenamefont {Lodahl}}]{Johansen2010}%
  \BibitemOpen
  \bibfield  {author} {\bibinfo {author} {\bibfnamefont {J.}~\bibnamefont
  {Johansen}}, \bibinfo {author} {\bibfnamefont {B.}~\bibnamefont {Julsgaard}},
  \bibinfo {author} {\bibfnamefont {S.}~\bibnamefont {Stobbe}}, \bibinfo
  {author} {\bibfnamefont {J.~M.}\ \bibnamefont {Hvam}}, \ and\ \bibinfo
  {author} {\bibfnamefont {P.}~\bibnamefont {Lodahl}},\ }\href@noop {}
  {\bibfield  {journal} {\bibinfo  {journal} {Phys. Rev. B}\ }\textbf {\bibinfo
  {volume} {81}},\ \bibinfo {pages} {081304} (\bibinfo {year}
  {2010})}\BibitemShut {NoStop}%
\bibitem [{\citenamefont {Davan\c{c}o}\ \emph {et~al.}(2014)\citenamefont
  {Davan\c{c}o}, \citenamefont {Hellberg}, \citenamefont {Ates}, \citenamefont
  {Badolato},\ and\ \citenamefont {Srinivasan}}]{Davanco2014}%
  \BibitemOpen
  \bibfield  {author} {\bibinfo {author} {\bibfnamefont {M.}~\bibnamefont
  {Davan\c{c}o}}, \bibinfo {author} {\bibfnamefont {C.~S.}\ \bibnamefont
  {Hellberg}}, \bibinfo {author} {\bibfnamefont {S.}~\bibnamefont {Ates}},
  \bibinfo {author} {\bibfnamefont {A.}~\bibnamefont {Badolato}}, \ and\
  \bibinfo {author} {\bibfnamefont {K.}~\bibnamefont {Srinivasan}},\
  }\href@noop {} {\bibfield  {journal} {\bibinfo  {journal} {Phys. Rev. B}\
  }\textbf {\bibinfo {volume} {89}},\ \bibinfo {pages} {161303} (\bibinfo
  {year} {2014})}\BibitemShut {NoStop}%
\bibitem [{\citenamefont {Zhou}\ \emph {et~al.}(2018)\citenamefont {Zhou},
  \citenamefont {Kulkova}, \citenamefont {Lund-Hansen}, \citenamefont {Hansen},
  \citenamefont {Lodahl},\ and\ \citenamefont {Midolo}}]{zhou2018}%
  \BibitemOpen
  \bibfield  {author} {\bibinfo {author} {\bibfnamefont {X.}~\bibnamefont
  {Zhou}}, \bibinfo {author} {\bibfnamefont {I.}~\bibnamefont {Kulkova}},
  \bibinfo {author} {\bibfnamefont {T.}~\bibnamefont {Lund-Hansen}}, \bibinfo
  {author} {\bibfnamefont {S.~L.}\ \bibnamefont {Hansen}}, \bibinfo {author}
  {\bibfnamefont {P.}~\bibnamefont {Lodahl}}, \ and\ \bibinfo {author}
  {\bibfnamefont {L.}~\bibnamefont {Midolo}},\ }\href@noop {} {\bibfield
  {journal} {\bibinfo  {journal} {Appl. Phys. Lett.}\ }\textbf {\bibinfo
  {volume} {113}},\ \bibinfo {pages} {251103} (\bibinfo {year}
  {2018})}\BibitemShut {NoStop}%
\bibitem [{\citenamefont {Scarpelli}\ \emph {et~al.}(2019)\citenamefont
  {Scarpelli}, \citenamefont {Lang}, \citenamefont {Masia}, \citenamefont
  {Beggs}, \citenamefont {Muljarov}, \citenamefont {Young}, \citenamefont
  {Oulton}, \citenamefont {Kamp}, \citenamefont {H\"ofling}, \citenamefont
  {Schneider},\ and\ \citenamefont {Langbein}}]{Scarpelli2019}%
  \BibitemOpen
  \bibfield  {author} {\bibinfo {author} {\bibfnamefont {L.}~\bibnamefont
  {Scarpelli}}, \bibinfo {author} {\bibfnamefont {B.}~\bibnamefont {Lang}},
  \bibinfo {author} {\bibfnamefont {F.}~\bibnamefont {Masia}}, \bibinfo
  {author} {\bibfnamefont {D.~M.}\ \bibnamefont {Beggs}}, \bibinfo {author}
  {\bibfnamefont {E.~A.}\ \bibnamefont {Muljarov}}, \bibinfo {author}
  {\bibfnamefont {A.~B.}\ \bibnamefont {Young}}, \bibinfo {author}
  {\bibfnamefont {R.}~\bibnamefont {Oulton}}, \bibinfo {author} {\bibfnamefont
  {M.}~\bibnamefont {Kamp}}, \bibinfo {author} {\bibfnamefont {S.}~\bibnamefont
  {H\"ofling}}, \bibinfo {author} {\bibfnamefont {C.}~\bibnamefont
  {Schneider}}, \ and\ \bibinfo {author} {\bibfnamefont {W.}~\bibnamefont
  {Langbein}},\ }\href@noop {} {\bibfield  {journal} {\bibinfo  {journal}
  {Phys. Rev. B}\ }\textbf {\bibinfo {volume} {100}},\ \bibinfo {pages}
  {035311} (\bibinfo {year} {2019})}\BibitemShut {NoStop}%
\bibitem [{\citenamefont {Pu}\ \emph {et~al.}(2010)\citenamefont {Pu},
  \citenamefont {Liu}, \citenamefont {Ou}, \citenamefont {Yvind},\ and\
  \citenamefont {Hvam}}]{Pu2010}%
  \BibitemOpen
  \bibfield  {author} {\bibinfo {author} {\bibfnamefont {M.}~\bibnamefont
  {Pu}}, \bibinfo {author} {\bibfnamefont {L.}~\bibnamefont {Liu}}, \bibinfo
  {author} {\bibfnamefont {H.}~\bibnamefont {Ou}}, \bibinfo {author}
  {\bibfnamefont {K.}~\bibnamefont {Yvind}}, \ and\ \bibinfo {author}
  {\bibfnamefont {J.~M.}\ \bibnamefont {Hvam}},\ }\href@noop {} {\bibfield
  {journal} {\bibinfo  {journal} {Opt. Commun.}\ }\textbf {\bibinfo {volume}
  {283}},\ \bibinfo {pages} {3678} (\bibinfo {year} {2010})}\BibitemShut
  {NoStop}%
\bibitem [{\citenamefont {Tiecke}\ \emph {et~al.}(2015)\citenamefont {Tiecke},
  \citenamefont {Nayak}, \citenamefont {Thompson}, \citenamefont {Peyronel},
  \citenamefont {de~Leon}, \citenamefont {Vuleti{\'c}},\ and\ \citenamefont
  {Lukin}}]{Tiecke2015}%
  \BibitemOpen
  \bibfield  {author} {\bibinfo {author} {\bibfnamefont {T.}~\bibnamefont
  {Tiecke}}, \bibinfo {author} {\bibfnamefont {K.}~\bibnamefont {Nayak}},
  \bibinfo {author} {\bibfnamefont {J.~D.}\ \bibnamefont {Thompson}}, \bibinfo
  {author} {\bibfnamefont {T.}~\bibnamefont {Peyronel}}, \bibinfo {author}
  {\bibfnamefont {N.~P.}\ \bibnamefont {de~Leon}}, \bibinfo {author}
  {\bibfnamefont {V.}~\bibnamefont {Vuleti{\'c}}}, \ and\ \bibinfo {author}
  {\bibfnamefont {M.}~\bibnamefont {Lukin}},\ }\href@noop {} {\bibfield
  {journal} {\bibinfo  {journal} {Optica}\ }\textbf {\bibinfo {volume} {2}},\
  \bibinfo {pages} {70} (\bibinfo {year} {2015})}\BibitemShut {NoStop}%
\bibitem [{\citenamefont {Aaronson}\ and\ \citenamefont
  {Arkhipov}(2013)}]{Aaronson2013}%
  \BibitemOpen
  \bibfield  {author} {\bibinfo {author} {\bibfnamefont {S.}~\bibnamefont
  {Aaronson}}\ and\ \bibinfo {author} {\bibfnamefont {A.}~\bibnamefont
  {Arkhipov}},\ }\href@noop {} {\bibfield  {journal} {\bibinfo  {journal}
  {Theor. Comput.}\ }\textbf {\bibinfo {volume} {9}},\ \bibinfo {pages} {143}
  (\bibinfo {year} {2013})}\BibitemShut {NoStop}%
\bibitem [{\citenamefont {Jerrum}\ \emph {et~al.}(2004)\citenamefont {Jerrum},
  \citenamefont {Sinclair},\ and\ \citenamefont {Vigoda}}]{jerrum2004}%
  \BibitemOpen
  \bibfield  {author} {\bibinfo {author} {\bibfnamefont {M.}~\bibnamefont
  {Jerrum}}, \bibinfo {author} {\bibfnamefont {A.}~\bibnamefont {Sinclair}}, \
  and\ \bibinfo {author} {\bibfnamefont {E.}~\bibnamefont {Vigoda}},\
  }\href@noop {} {\bibfield  {journal} {\bibinfo  {journal} {J. ACM}\ }\textbf
  {\bibinfo {volume} {51}},\ \bibinfo {pages} {671} (\bibinfo {year}
  {2004})}\BibitemShut {NoStop}%
\bibitem [{\citenamefont {Arkhipov}(2015)}]{Arkhipov2015}%
  \BibitemOpen
  \bibfield  {author} {\bibinfo {author} {\bibfnamefont {A.}~\bibnamefont
  {Arkhipov}},\ }\href@noop {} {\bibfield  {journal} {\bibinfo  {journal}
  {Phys. Rev. A}\ }\textbf {\bibinfo {volume} {92}},\ \bibinfo {pages} {062326}
  (\bibinfo {year} {2015})}\BibitemShut {NoStop}%
\bibitem [{\citenamefont {Shchesnovich}(2014)}]{Shchesnovich2014}%
  \BibitemOpen
  \bibfield  {author} {\bibinfo {author} {\bibfnamefont {V.~S.}\ \bibnamefont
  {Shchesnovich}},\ }\href@noop {} {\bibfield  {journal} {\bibinfo  {journal}
  {Phys. Rev. A}\ }\textbf {\bibinfo {volume} {89}},\ \bibinfo {pages} {022333}
  (\bibinfo {year} {2014})}\BibitemShut {NoStop}%
\bibitem [{\citenamefont {Rohde}(2015)}]{Rhode2015}%
  \BibitemOpen
  \bibfield  {author} {\bibinfo {author} {\bibfnamefont {P.~P.}\ \bibnamefont
  {Rohde}},\ }\href@noop {} {\bibfield  {journal} {\bibinfo  {journal} {Phys.
  Rev. A}\ }\textbf {\bibinfo {volume} {91}},\ \bibinfo {pages} {012307}
  (\bibinfo {year} {2015})}\BibitemShut {NoStop}%
\bibitem [{\citenamefont {Wang}\ \emph {et~al.}(2018)\citenamefont {Wang},
  \citenamefont {Li}, \citenamefont {Jiang}, \citenamefont {He}, \citenamefont
  {Li}, \citenamefont {Ding}, \citenamefont {Chen}, \citenamefont {Qin},
  \citenamefont {Peng}, \citenamefont {Schneider}, \citenamefont {Kamp},
  \citenamefont {Zhang}, \citenamefont {Li}, \citenamefont {You}, \citenamefont
  {Wang}, \citenamefont {Dowling}, \citenamefont {H\"ofling}, \citenamefont
  {Lu},\ and\ \citenamefont {Pan}}]{Wang2018}%
  \BibitemOpen
  \bibfield  {author} {\bibinfo {author} {\bibfnamefont {H.}~\bibnamefont
  {Wang}}, \bibinfo {author} {\bibfnamefont {W.}~\bibnamefont {Li}}, \bibinfo
  {author} {\bibfnamefont {X.}~\bibnamefont {Jiang}}, \bibinfo {author}
  {\bibfnamefont {Y.-M.}\ \bibnamefont {He}}, \bibinfo {author} {\bibfnamefont
  {Y.-H.}\ \bibnamefont {Li}}, \bibinfo {author} {\bibfnamefont
  {X.}~\bibnamefont {Ding}}, \bibinfo {author} {\bibfnamefont {M.-C.}\
  \bibnamefont {Chen}}, \bibinfo {author} {\bibfnamefont {J.}~\bibnamefont
  {Qin}}, \bibinfo {author} {\bibfnamefont {C.-Z.}\ \bibnamefont {Peng}},
  \bibinfo {author} {\bibfnamefont {C.}~\bibnamefont {Schneider}}, \bibinfo
  {author} {\bibfnamefont {M.}~\bibnamefont {Kamp}}, \bibinfo {author}
  {\bibfnamefont {W.-J.}\ \bibnamefont {Zhang}}, \bibinfo {author}
  {\bibfnamefont {H.}~\bibnamefont {Li}}, \bibinfo {author} {\bibfnamefont
  {L.-X.}\ \bibnamefont {You}}, \bibinfo {author} {\bibfnamefont
  {Z.}~\bibnamefont {Wang}}, \bibinfo {author} {\bibfnamefont {J.~P.}\
  \bibnamefont {Dowling}}, \bibinfo {author} {\bibfnamefont {S.}~\bibnamefont
  {H\"ofling}}, \bibinfo {author} {\bibfnamefont {C.-Y.}\ \bibnamefont {Lu}}, \
  and\ \bibinfo {author} {\bibfnamefont {J.-W.}\ \bibnamefont {Pan}},\
  }\href@noop {} {\bibfield  {journal} {\bibinfo  {journal} {Phys. Rev. Lett.}\
  }\textbf {\bibinfo {volume} {120}},\ \bibinfo {pages} {230502} (\bibinfo
  {year} {2018})}\BibitemShut {NoStop}%
\bibitem [{\citenamefont {Rahimi-Keshari}\ \emph {et~al.}(2016)\citenamefont
  {Rahimi-Keshari}, \citenamefont {Ralph},\ and\ \citenamefont
  {Caves}}]{Keshari2016}%
  \BibitemOpen
  \bibfield  {author} {\bibinfo {author} {\bibfnamefont {S.}~\bibnamefont
  {Rahimi-Keshari}}, \bibinfo {author} {\bibfnamefont {T.~C.}\ \bibnamefont
  {Ralph}}, \ and\ \bibinfo {author} {\bibfnamefont {C.~M.}\ \bibnamefont
  {Caves}},\ }\href@noop {} {\bibfield  {journal} {\bibinfo  {journal} {Phys.
  Rev. X}\ }\textbf {\bibinfo {volume} {6}},\ \bibinfo {pages} {021039}
  (\bibinfo {year} {2016})}\BibitemShut {NoStop}%
\bibitem [{\citenamefont {Neville}\ \emph {et~al.}(2017)\citenamefont
  {Neville}, \citenamefont {Sparrow}, \citenamefont {Clifford}, \citenamefont
  {Johnston}, \citenamefont {Birchall}, \citenamefont {Montanaro},\ and\
  \citenamefont {Laing}}]{Neville2017}%
  \BibitemOpen
  \bibfield  {author} {\bibinfo {author} {\bibfnamefont {A.}~\bibnamefont
  {Neville}}, \bibinfo {author} {\bibfnamefont {C.}~\bibnamefont {Sparrow}},
  \bibinfo {author} {\bibfnamefont {R.}~\bibnamefont {Clifford}}, \bibinfo
  {author} {\bibfnamefont {E.}~\bibnamefont {Johnston}}, \bibinfo {author}
  {\bibfnamefont {P.~M.}\ \bibnamefont {Birchall}}, \bibinfo {author}
  {\bibfnamefont {A.}~\bibnamefont {Montanaro}}, \ and\ \bibinfo {author}
  {\bibfnamefont {A.}~\bibnamefont {Laing}},\ }\href@noop {} {\bibfield
  {journal} {\bibinfo  {journal} {Nat. Phys.}\ }\textbf {\bibinfo {volume}
  {13}},\ \bibinfo {pages} {1153} (\bibinfo {year} {2017})}\BibitemShut
  {NoStop}%
\bibitem [{\citenamefont {Clifford}\ and\ \citenamefont
  {Clifford}(2018)}]{Clifford2018}%
  \BibitemOpen
  \bibfield  {author} {\bibinfo {author} {\bibfnamefont {P.}~\bibnamefont
  {Clifford}}\ and\ \bibinfo {author} {\bibfnamefont {R.}~\bibnamefont
  {Clifford}},\ }in\ \href@noop {} {\emph {\bibinfo {booktitle} {Proceedings of
  the Twenty-Ninth Annual ACM-SIAM Symposium on Discrete Algorithms}}}\
  (\bibinfo {organization} {SIAM},\ \bibinfo {year} {2018})\ pp.\ \bibinfo
  {pages} {146--155}\BibitemShut {NoStop}%
\bibitem [{\citenamefont {Oszmaniec}\ and\ \citenamefont
  {Brod}(2018)}]{Oszmaniec_2018}%
  \BibitemOpen
  \bibfield  {author} {\bibinfo {author} {\bibfnamefont {M.}~\bibnamefont
  {Oszmaniec}}\ and\ \bibinfo {author} {\bibfnamefont {D.~J.}\ \bibnamefont
  {Brod}},\ }\href@noop {} {\bibfield  {journal} {\bibinfo  {journal} {New J.
  Phys.}\ }\textbf {\bibinfo {volume} {20}},\ \bibinfo {pages} {092002}
  (\bibinfo {year} {2018})}\BibitemShut {NoStop}%
\bibitem [{\citenamefont {Garc\'{i}a-Patr\'{o}n}\ \emph
  {et~al.}(2019)\citenamefont {Garc\'{i}a-Patr\'{o}n}, \citenamefont {Renema},\
  and\ \citenamefont {Shchesnovich}}]{Patron2019}%
  \BibitemOpen
  \bibfield  {author} {\bibinfo {author} {\bibfnamefont {R.}~\bibnamefont
  {Garc\'{i}a-Patr\'{o}n}}, \bibinfo {author} {\bibfnamefont {J.~J.}\
  \bibnamefont {Renema}}, \ and\ \bibinfo {author} {\bibfnamefont {V.~S.}\
  \bibnamefont {Shchesnovich}},\ }\href@noop {} {\bibfield  {journal} {\bibinfo
   {journal} {Quantum}\ }\textbf {\bibinfo {volume} {3}},\ \bibinfo {pages}
  {169} (\bibinfo {year} {2019})}\BibitemShut {NoStop}%
\bibitem [{\citenamefont {Moylett}\ \emph {et~al.}(2020)\citenamefont
  {Moylett}, \citenamefont {Garc\'{i}a-Patr\'{o}n}, \citenamefont {Renema},\
  and\ \citenamefont {Turner}}]{Moylett2020}%
  \BibitemOpen
  \bibfield  {author} {\bibinfo {author} {\bibfnamefont {A.~E.}\ \bibnamefont
  {Moylett}}, \bibinfo {author} {\bibfnamefont {R.}~\bibnamefont
  {Garc\'{i}a-Patr\'{o}n}}, \bibinfo {author} {\bibfnamefont {J.~J.}\
  \bibnamefont {Renema}}, \ and\ \bibinfo {author} {\bibfnamefont {P.~S.}\
  \bibnamefont {Turner}},\ }\href@noop {} {\bibfield  {journal} {\bibinfo
  {journal} {Quantum Sci. Technol.}\ }\textbf {\bibinfo {volume} {5}},\
  \bibinfo {pages} {015001} (\bibinfo {year} {2020})}\BibitemShut {NoStop}%
\bibitem [{Sum(2019)}]{Summit2019}%
  \BibitemOpen
  \href@noop {} {\enquote {\bibinfo {title} {{Introducing Summit}},}\ }\bibinfo
  {howpublished} {\url{https://www.olcf.ornl.gov/summit/}} (\bibinfo {year}
  {2019}),\ \bibinfo {note} {accessed: 2020-03-04}\BibitemShut {NoStop}%
\bibitem [{\citenamefont {Ferrante}\ and\ \citenamefont
  {Frigo}(2012)}]{Ferrante2012}%
  \BibitemOpen
  \bibfield  {author} {\bibinfo {author} {\bibfnamefont {M.}~\bibnamefont
  {Ferrante}}\ and\ \bibinfo {author} {\bibfnamefont {N.}~\bibnamefont
  {Frigo}},\ }\href@noop {} {\bibfield  {journal} {\bibinfo  {journal}
  {arXiv:1209.2667}\ } (\bibinfo {year} {2012})}\BibitemShut {NoStop}%
\bibitem [{\citenamefont {Esmaeil~Zadeh}\ \emph {et~al.}(2017)\citenamefont
  {Esmaeil~Zadeh}, \citenamefont {Los}, \citenamefont {Gourgues}, \citenamefont
  {Steinmetz}, \citenamefont {Bulgarini}, \citenamefont {Dobrovolskiy},
  \citenamefont {Zwiller},\ and\ \citenamefont {Dorenbos}}]{zwiller2017}%
  \BibitemOpen
  \bibfield  {author} {\bibinfo {author} {\bibfnamefont {I.}~\bibnamefont
  {Esmaeil~Zadeh}}, \bibinfo {author} {\bibfnamefont {J.~W.}\ \bibnamefont
  {Los}}, \bibinfo {author} {\bibfnamefont {R.~B.}\ \bibnamefont {Gourgues}},
  \bibinfo {author} {\bibfnamefont {V.}~\bibnamefont {Steinmetz}}, \bibinfo
  {author} {\bibfnamefont {G.}~\bibnamefont {Bulgarini}}, \bibinfo {author}
  {\bibfnamefont {S.~M.}\ \bibnamefont {Dobrovolskiy}}, \bibinfo {author}
  {\bibfnamefont {V.}~\bibnamefont {Zwiller}}, \ and\ \bibinfo {author}
  {\bibfnamefont {S.~N.}\ \bibnamefont {Dorenbos}},\ }\href@noop {} {\bibfield
  {journal} {\bibinfo  {journal} {APL Photonics}\ }\textbf {\bibinfo {volume}
  {2}},\ \bibinfo {pages} {111301} (\bibinfo {year} {2017})}\BibitemShut
  {NoStop}%
\bibitem [{\citenamefont {Arkhipov}\ and\ \citenamefont
  {Kuperberg}(2012)}]{Arkhipov2012}%
  \BibitemOpen
  \bibfield  {author} {\bibinfo {author} {\bibfnamefont {A.}~\bibnamefont
  {Arkhipov}}\ and\ \bibinfo {author} {\bibfnamefont {G.}~\bibnamefont
  {Kuperberg}},\ }\href@noop {} {\bibfield  {journal} {\bibinfo  {journal}
  {Geom. Topol. Monog.}\ }\textbf {\bibinfo {volume} {18}},\ \bibinfo {pages}
  {1} (\bibinfo {year} {2012})}\BibitemShut {NoStop}%
\bibitem [{\citenamefont {Spagnolo}\ \emph {et~al.}(2014)\citenamefont
  {Spagnolo}, \citenamefont {Vitelli}, \citenamefont {Bentivegna},
  \citenamefont {Brod}, \citenamefont {Crespi}, \citenamefont {Flamini},
  \citenamefont {Giacomini}, \citenamefont {Milani}, \citenamefont {Ramponi},
  \citenamefont {Mataloni}, \citenamefont {Osellame}, \citenamefont {{a}om
  E.~F.},\ and\ \citenamefont {Sciarrino}}]{Spagnolo2014}%
  \BibitemOpen
  \bibfield  {author} {\bibinfo {author} {\bibfnamefont {N.}~\bibnamefont
  {Spagnolo}}, \bibinfo {author} {\bibfnamefont {C.}~\bibnamefont {Vitelli}},
  \bibinfo {author} {\bibfnamefont {M.}~\bibnamefont {Bentivegna}}, \bibinfo
  {author} {\bibfnamefont {D.~J.}\ \bibnamefont {Brod}}, \bibinfo {author}
  {\bibfnamefont {A.}~\bibnamefont {Crespi}}, \bibinfo {author} {\bibfnamefont
  {F.}~\bibnamefont {Flamini}}, \bibinfo {author} {\bibfnamefont
  {S.}~\bibnamefont {Giacomini}}, \bibinfo {author} {\bibfnamefont
  {G.}~\bibnamefont {Milani}}, \bibinfo {author} {\bibfnamefont
  {R.}~\bibnamefont {Ramponi}}, \bibinfo {author} {\bibfnamefont
  {P.}~\bibnamefont {Mataloni}}, \bibinfo {author} {\bibfnamefont
  {R.}~\bibnamefont {Osellame}}, \bibinfo {author} {\bibfnamefont
  {G.}~\bibnamefont {{a}om E.~F.}}, \ and\ \bibinfo {author} {\bibfnamefont
  {F.}~\bibnamefont {Sciarrino}},\ }\href@noop {} {\bibfield  {journal}
  {\bibinfo  {journal} {Nat. Photon.}\ }\textbf {\bibinfo {volume} {8}},\
  \bibinfo {pages} {615} (\bibinfo {year} {2014})}\BibitemShut {NoStop}%
\bibitem [{\citenamefont {Carolan}\ \emph {et~al.}(2014)\citenamefont
  {Carolan}, \citenamefont {Meinecke}, \citenamefont {Shadbolt}, \citenamefont
  {Russell}, \citenamefont {Ismail}, \citenamefont {W\"{o}rhoff}, \citenamefont
  {Rudolph}, \citenamefont {Thompson}, \citenamefont {O'Brien}, \citenamefont
  {Matthews},\ and\ \citenamefont {Laing}}]{Carolan2014}%
  \BibitemOpen
  \bibfield  {author} {\bibinfo {author} {\bibfnamefont {J.}~\bibnamefont
  {Carolan}}, \bibinfo {author} {\bibfnamefont {J.~D.~A.}\ \bibnamefont
  {Meinecke}}, \bibinfo {author} {\bibfnamefont {P.~J.}\ \bibnamefont
  {Shadbolt}}, \bibinfo {author} {\bibfnamefont {N.~J.}\ \bibnamefont
  {Russell}}, \bibinfo {author} {\bibfnamefont {N.}~\bibnamefont {Ismail}},
  \bibinfo {author} {\bibfnamefont {K.}~\bibnamefont {W\"{o}rhoff}}, \bibinfo
  {author} {\bibfnamefont {T.}~\bibnamefont {Rudolph}}, \bibinfo {author}
  {\bibfnamefont {M.~G.}\ \bibnamefont {Thompson}}, \bibinfo {author}
  {\bibfnamefont {J.~L.}\ \bibnamefont {O'Brien}}, \bibinfo {author}
  {\bibfnamefont {J.~C.~F.}\ \bibnamefont {Matthews}}, \ and\ \bibinfo {author}
  {\bibfnamefont {A.}~\bibnamefont {Laing}},\ }\href@noop {} {\bibfield
  {journal} {\bibinfo  {journal} {Nat. Photon.}\ }\textbf {\bibinfo {volume}
  {8}},\ \bibinfo {pages} {621} (\bibinfo {year} {2014})}\BibitemShut {NoStop}%
\bibitem [{\citenamefont {Liu}\ \emph {et~al.}(2016)\citenamefont {Liu},
  \citenamefont {Lund}, \citenamefont {Gu},\ and\ \citenamefont
  {Ralph}}]{Liu2016}%
  \BibitemOpen
  \bibfield  {author} {\bibinfo {author} {\bibfnamefont {K.}~\bibnamefont
  {Liu}}, \bibinfo {author} {\bibfnamefont {A.~P.}\ \bibnamefont {Lund}},
  \bibinfo {author} {\bibfnamefont {Y.-J.}\ \bibnamefont {Gu}}, \ and\ \bibinfo
  {author} {\bibfnamefont {T.~C.}\ \bibnamefont {Ralph}},\ }\href@noop {}
  {\bibfield  {journal} {\bibinfo  {journal} {J. Opt. Soc. Am. B}\ }\textbf
  {\bibinfo {volume} {33}},\ \bibinfo {pages} {1835} (\bibinfo {year}
  {2016})}\BibitemShut {NoStop}%
\bibitem [{\citenamefont {Agresti}\ \emph {et~al.}(2019)\citenamefont
  {Agresti}, \citenamefont {Viggianiello}, \citenamefont {Flamini},
  \citenamefont {Spagnolo}, \citenamefont {Crespi}, \citenamefont {Osellame},
  \citenamefont {Wiebe},\ and\ \citenamefont {Sciarrino}}]{Agresti2019}%
  \BibitemOpen
  \bibfield  {author} {\bibinfo {author} {\bibfnamefont {I.}~\bibnamefont
  {Agresti}}, \bibinfo {author} {\bibfnamefont {N.}~\bibnamefont
  {Viggianiello}}, \bibinfo {author} {\bibfnamefont {F.}~\bibnamefont
  {Flamini}}, \bibinfo {author} {\bibfnamefont {N.}~\bibnamefont {Spagnolo}},
  \bibinfo {author} {\bibfnamefont {A.}~\bibnamefont {Crespi}}, \bibinfo
  {author} {\bibfnamefont {R.}~\bibnamefont {Osellame}}, \bibinfo {author}
  {\bibfnamefont {N.}~\bibnamefont {Wiebe}}, \ and\ \bibinfo {author}
  {\bibfnamefont {F.}~\bibnamefont {Sciarrino}},\ }\href@noop {} {\bibfield
  {journal} {\bibinfo  {journal} {Phys. Rev. X}\ }\textbf {\bibinfo {volume}
  {9}},\ \bibinfo {pages} {011013} (\bibinfo {year} {2019})}\BibitemShut
  {NoStop}%
\end{thebibliography}
\end{document}